\documentclass{new_tlp}
\newcommand\citet[1]{\citeauthor{#1}~[\citeyear{#1}]}
\usepackage{mathptmx}

\usepackage{times}

\usepackage{amssymb}

\usepackage{graphicx}
\usepackage{paralist}

\usepackage[linesnumbered]{algorithm2e}

\usepackage{url}
\usepackage{import}

\usepackage{ifthen}
\usepackage{url}
\usepackage{amssymb}
\usepackage{amsmath}
\usepackage{import}
\usepackage{xparse}
\usepackage{amsmath}
\usepackage[usenames,dvipsnames]{xcolor}
\usepackage{xspace}
\usepackage{datatool}
\usepackage{glossaries}

\providecommand\m[1]{\ensuremath{{#1}}\xspace}
\renewcommand{\m}[1]{\ensuremath{{#1}}\xspace}

	\newcommand{\limplies}{\Rightarrow}

	\newcommand{\lrule}{\leftarrow}
	\newcommand{\cause}{\stackrel{c}{\lrule}}

	\newcommand{\voc}{\m{\Sigma}}

	\newcommand{\struct}{\m{I}}

	\newcommand{\theory}{\m{\mathcal{T}}}

	\newcommand{\f}{\m{\varphi}}

	\NewDocumentCommand\inter{g+g}{%
	  \IfNoValueTF{#1}
	    {\struct}
	    {\m{#1^{#2}}}}

	\newcommand{\ttt}{\m{\overline{t}}}

	\newcommand{\bool}{\m{\mathbb{B}}}

	\renewcommand{\int}{\m{\mathbb{Z}}}

	\NewDocumentCommand\subs{g+g}{%
	  \IfNoValueTF{#1}
	    {\m{/}}
	    {\m{#1/ #2}}}

	\newcommand{\logicname}[1]{\textsc{#1}\xspace}

	\newcommand{\idp}{\logicname{IDP}}

	\newcommand{\saucy}{\logicname{Saucy}}
	\newcommand{\sbass}{\logicname{sbass}}

	\newcommand{\gringo}{\logicname{gringo}}

	\newcommand{\fodot}{\logicname{FO(\ensuremath{\cdot})}}

\newcommand{\ouracronym}[3]{%
	\newacronym{#1}{#2}{#3}
	\expandafter\newcommand\csname #1\endcsname{\gls{#1}\xspace}%
}
	\ouracronym{FO}{FO}{first-order logic}
	\ouracronym{PC}{PC}{propositional calculus}
	\ouracronym{MX}{MX}{Model Expansion}
	\ouracronym{MO}{MO}{Model Optimization}
	\ouracronym{ASP}{ASP}{Answer Set Programming}
	\ouracronym{TP}{TP}{Theorem Proving}
	\ouracronym{LP}{LP}{Logic Programming}
	\ouracronym{CP}{CP}{Constraint Programming}
	\ouracronym{FP}{FP}{Functional Programming}
	\ouracronym{KR}{KR}{Knowledge Representation}
	\ouracronym{CSP}{CSP}{Constraint Satisfaction Problem}
	\ouracronym{SMT}{SMT}{SAT Modulo Theories}
	\ouracronym{KBS}{KBS}{Knowledge Base System}
	\ouracronym{NNF}{NNF}{Negation Normal Form}
	\ouracronym{FNNF}{FNNF}{Flat Negation Normal Form}
	\ouracronym{DefNNF}{DefNNF}{Definition Negation Normal Form}
	\ouracronym{DEFNF}{DEFNF}{Definition Normal Form}
	\ouracronym{CDCL}{CDCL}{Conflict-Driven Clause-Learning}
	\ouracronym{WFS}{WFS}{Well-Founded Semantics}
	\ouracronym{LCG}{LCG}{Lazy Clause Generation}
	\ouracronym{AEL}{AEL}{Autoepistemic Logic}
	\ouracronym{OEL}{OEL}{Ordered Epistemic Logic}
	\ouracronym{AFT}{AFT}{Approximation Fixpoint Theory}

	\makeatletter
	\def\ifenv#1{
	\def\@tempa{#1}%
	\def\@ttempa{#1*}%
	\ifx\@tempa\@currenvir
	\expandafter\@firstoftwo
	\else
	\expandafter\@secondoftwo
	\fi
	}
	\makeatother

	\newcommand{\ddrule}[4]{\ensuremath{#1 \leftarrow #2 & \{#3\} & #4}}
	\newcommand{\drule}[2]{\ensuremath{#1 & \leftarrow & #2}}

	\newcommand{\darule}[4]{\ensuremath{#1 \leftarrow #2 & \{#3\} & #4}}
	\newcommand{\arule}[2]{\ensuremath{#1 \, &\leftarrow \, #2}}

	\newenvironment{ltheo}{\[\begin{array}{l}}{\end{array}\]\ignorespacesafterend}

	\newcommand{\LNDRule}[2]{
	\ifenv{array}
	{\drule{#1}{#2}}
	{ \ifenv{align}
		{\arule{#1}{#2}}
		{\ifenv{align*}
		{\arule{#1}{#2}}
		{ERROR: using LDRule in unsupported environment: \@currenvir}
		}
	}
	}

	\newcommand{\LDRule}[4]{
	\ifenv{array}
	{\ddrule{#1}{#2}{#3}{#4}}
	{ \ifenv{align}
		{\darule{#1}{#2}{#3}{#4}}
		{\ifenv{align*}
		{\darule{#1}{#2}{#3}{#4}}
		{ERROR: using LDRule in unsupported environment: \@currenvir}
		}
	}
	}

	\NewDocumentCommand\LRule{m+g+g+g}{%
		\IfNoValueTF{#2}%
		{#1.&}{%
		\IfNoValueTF{#3}
		{\LNDRule{#1}{#2.}}
		{\LDRule{#1}{#2.}{#3}{#4}}%
		}
	}

	\NewDocumentCommand\CLRule{m+g}{%
	\ifenv{array}
	{\cdrule{#1}{#2}}
	{ \ifenv{align}
		{\carule{#1}{#2}}
		{\ifenv{align*}
			{\carule{#1}{#2}}
			{ERROR: using CLRule in unsupported environment: \@currenvir}
		}
	}
	}

	\NewDocumentCommand\carule{m+g}{%
		\IfNoValueTF{#2}
			{\ensuremath{#1.}}
			{\ensuremath{#1 \, &\cause \, #2}}}
	\NewDocumentCommand\cdrule{m+g}{%
		\IfNoValueTF{#2}
			{\ensuremath{#1.}}
			{\ensuremath{#1 & \cause & #2}}}

	\newcommand{\algrule}[4]{
	\hbox{{#1}:}& 
	\quad #2 ~\longrightarrow~ #3 
	\hbox{~ if } #4\\
	}

	\newcommand{\AlgoRule}[4]{
	\ifenv{array}
	{\algrule{#1}{#2}{#3}{#4}}
		{ERROR: using AlgoRule in unsupported environment: \@currenvir}
	}

\newcommand{\commentstyle}{\color{Gray}}

	\usepackage[final]{listings}

	\lstdefinelanguage{idp}{
		morekeywords=[1]{query(}, 
		morekeywords=[2]{namespace,vocabulary,theory,structure,procedure,term,set,formula, spec, specification,query},
		morekeywords=[3]{include,using,type,isa,contains,partial,extern,LFD,GFD,constructed,from,constraint,pred,supertype,of,subtype,define},
		morekeywords=[4]{int,float,char,string,nat},
		morekeywords=[5]{if,then,else,for,end},
		morecomment=[s]{/*}{*/},	
		morecomment=[l]{//}
	}
	\newcommand{\ignore}[1]{}

	\newboolean{nocomments}
	\setboolean{nocomments}{false}

	\newboolean{commentmargin}
	\setboolean{commentmargin}{true}

	\newcommand{\namedcomment}[3]{
		\ifthenelse{\boolean{nocomments}}
		{} 
		{ 
			\ifthenelse{\boolean{commentmargin}}
				{ {\color{#3} \marginpar{\color{#3}\sc #2}#1}  } 
				{  {\color{#3} {\sc #2}: #1}  } 
		}
	}
	\newcommand{\mnamedcomment}[3]{\ifthenelse{\boolean{nocomments}}{}{{\marginpar{ \color{#3}{\sc #2}:#1}}}}

	\newcommand{\bart}[1]{\namedcomment{#1}{bb}{red}}

	\newcommand{\maurice}[1]{\namedcomment{#1}{mb}{orange}}

	\newcommand{\jo}[1]{ \namedcomment{#1}{jo}{Fuchsia}}

	\usepackage{soul}

\usepackage{etoolbox}

\newcommand\setcitation[2]{%
  \csdef{mycommoncitation#1}{#2}}
\newcommand\getcitation[1]{%
  \csuse{mycommoncitation#1}}

\setcitation{IDP}{WarrenBook/DeCatBBD14}
\setcitation{idp}{WarrenBook/DeCatBBD14}
\setcitation{fodot}{tocl/DeneckerT08}
\setcitation{CP}{Apt03}
\setcitation{cp}{Apt03}
\setcitation{EZCSP}{lpnmr/Balduccini11}
\setcitation{KR}{Baral:2003}
\setcitation{ASPComp2}{lpnmr/DeneckerVBGT09}
\setcitation{ASPComp3}{journals/tplp/CalimeriIR14}
\setcitation{ASPComp4}{conf/lpnmr/AlvianoCCDDIKKOPPRRSSSWX13}
\setcitation{CPSupport}{ictai/DeCat13}
\setcitation{CPsupport}{ictai/DeCat13}
\setcitation{functionDetection}{iclp/DeCatB13}
\setcitation{fodot2asp}{DeneckerLTV12} 
\setcitation{Tarskian}{DeneckerLTV12} 
\setcitation{Inca}{iclp/DrescherW12}
\setcitation{csp2asp}{ijcai/DrescherW11}
\setcitation{DPLLT}{cav/GanzingerHNOT04}
\setcitation{AspInPractice}{synthesis/2012Gebser}
\setcitation{clasp}{ai/GebserKS12}
\setcitation{oclingo}{kr/GebserGKOSS12}
\setcitation{gringo}{lpnmr/GebserST07}
\setcitation{cmodels}{aaai/GiunchigliaLM04}
\setcitation{inputster}{tplp/Jansen13}
\setcitation{DLV}{tocl/LeonePFEGPS06}
\setcitation{LearningPaper}{TPLP/BruynoogheBBDDJLRDV} 
\setcitation{clog}{iclp/BogaertsVDV14} 
\setcitation{foc}{iclp/BogaertsVDV14}
\setcitation{inferenceClog}{ecai/BogaertsVDV14}
\setcitation{examplesClog}{nmr/BogaertsVDV14b} 
\setcitation{AFT}{DeneckerBV12}
\setcitation{KBS}{iclp/DeneckerV08}
\setcitation{KBPE}{inap/DePooterWD11}
\setcitation{lazyGrounding}{jair/CatDBS15} 
\setcitation{lazygrounding}{jair/CatDBS15} 
\setcitation{ASP}{marek99stable}
\setcitation{satid}{sat/MarienWDB08}
\setcitation{lazyclausegeneration}{constraints/OhrimenkoSC09}
\setcitation{FP}{ACMCS/Hudak89}
\setcitation{GroundingWithBounds}{jair/WittocxMD10}
\setcitation{SAT}{faia/SilvaLM09}
\setcitation{LTC}{iclp/Bogaerts14}
\setcitation{SPSAT}{ictai/DevriendtBMDD12}
\setcitation{BreakID}{sat/DevriendtBBD16}
\setcitation{breakid}{sat/DevriendtBBD16}
\setcitation{LCG}{stuckeyLCG}
\setcitation{MiniZinc}{conf/cp/NethercoteSBBDT07}
\setcitation{minizinc}{conf/cp/NethercoteSBBDT07}
\setcitation{amadini}{cpaior/AmadiniGM13}
\setcitation{bootstrapping}{lash/BogaertsJDJBD14}
\setcitation{GroundedFixpoints}{ai/BogaertsVD15}
\setcitation{LogicBlox}{datalog/GreenAK12}
\setcitation{proB}{journals/sttt/LeuschelB08}
\setcitation{NaturalInductions}{KR/DeneckerV14} 
\setcitation{LP}{jacm/EmdenK76}
\setcitation{SMT}{faia/BarrettSST09}
\setcitation{AF}{ai/Dung95}
\setcitation{ADF}{kr/BrewkaW10}
\setcitation{af}{ai/Dung95}
\setcitation{adf}{kr/BrewkaW10}
\setcitation{ADFRevisited}{ijcai/BrewkaSEWW13}
\setcitation{adfrevisited}{ijcai/BrewkaSEWW13}
\setcitation{DefaultLogic}{ai/Reiter80}
\setcitation{DL}{ai/Reiter80}
\setcitation{AEL}{mo85}
\setcitation{minisat}{sat/EenS03}
\setcitation{completion}{adbt/Clark78}
\setcitation{wasp}{lpnmr/AlvianoDFLR13}
\setcitation{minisatid}{ictai/DeCat13}
\setcitation{lcg}{stuckeyLCG}
\setcitation{CEGAR}{jacm/ClarkeGJLV03}
\setcitation{cegar}{jacm/ClarkeGJLV03}
\setcitation{CuttingPlane}{or/DantzigFJ54}
\setcitation{kodkod}{tacas/TorlakJ07}
\setcitation{cdcl}{Marques-SilvaS99}
\setcitation{CDCL}{Marques-SilvaS99}
\setcitation{relevance}{ijcai/JansenBDJD16}
\setcitation{WFS}{GelderRS91}
\setcitation{wfs}{GelderRS91}
\setcitation{UnfoundedSet}{GelderRS91}
\setcitation{stablesemantics}{iclp/GelfondL88}
\setcitation{StableSemantics}{iclp/GelfondL88}
\setcitation{shatter}{Shatter}
\setcitation{sbass}{drtiwa11a}
\setcitation{lparsemanual}{url:lparse_manual}
\setcitation{}{}
\setcitation{}{}
\setcitation{}{}
\setcitation{}{}
\setcitation{}{}
\setcitation{}{}
\setcitation{}{}
\setcitation{}{}
\setcitation{}{}
\setcitation{}{}
\setcitation{}{}
\setcitation{}{}
\setcitation{}{}
\setcitation{}{}
\setcitation{}{}
\setcitation{}{}
\setcitation{}{}
\setcitation{}{}
\setcitation{}{}
\setcitation{}{}
\setcitation{}{}
\setcitation{}{}
\setcitation{}{}
\setcitation{}{}
\setcitation{}{}
\setcitation{}{}
\setcitation{}{}
\setcitation{}{}
\setcitation{}{}
\setcitation{}{}
  
\newcommand\refto[1]{%
      \getcitation{#1}}
      
\newcommand\mycite[1]{%
      \ifcsname mycommoncitation#1\endcsname%
   \cite{\getcitation{#1}}%
  \else%
    \cite{#1}%
  \fi%
}	
  
\ignore{

}

\usepackage{verbatim} 

\usepackage{tikz}
\usetikzlibrary{shapes,snakes}

\usepackage{scrextend}

\begin{document}

\title{On Local Domain Symmetry for Model Expansion}

\author[Jo Devriendt et al.]
       {Jo Devriendt$^\dagger$, Bart Bogaerts$^{\ddagger,\dagger}$, Maurice Bruynooghe$^\dagger$, Marc Denecker$^\dagger$ \\
    $^\dagger$KU Leuven -- University of Leuven, Celestijnenlaan 200A, Leuven, Belgium \\
    \email{firstname.lastname@cs.kuleuven.be}\\
    $^\ddagger$
    Helsinki Institute for Information Technology HIIT, Aalto University, FI-00076 AALTO, Finland}

%


\maketitle

\newcommand{\domain}{\m{D}}
\newcommand{\strucs}{\m{\Gamma_\domain}}
\newcommand{\strucsextending}{\m{\Gamma^{\struct}}}
\newcommand\symb[1]{\m{symb(#1)}}

\newcommand\argpos[2]{\m{#1|#2}}
\newcommand\localperm[3]{\tau^{#1}_{\argpos{#2}{#3}}}

\newcommand\domainof[1]{\m{\mathit{dom}(#1)}}
\newcommand\lex[1]{\m{\mathit{lex}^{\preceq_\domain}(#1)}}

\renewcommand\theory{\m{\mathnormal{T}}}

\newcommand\vocin{\m{\voc_{in}}}
\newcommand\vocout{\m{\voc_{out}}}
\newcommand\vocgc{\m{\voc_{gc}}}
\newcommand\vocgcin{\m{\voc_{gcin}}}
\newcommand\vocgcout{\m{\voc_{gcout}}}
\newcommand\vocstar{\m{\voc^*}}
\newcommand\vocinstar{\m{\voc_{in}^*}}
\newcommand\vocgcinstar{\m{\voc_{gcin}^*}}

\newcommand\theoryin{\m{\theory_{in}}}
\newcommand\theorygc{\m{\theory_{gc}}}
\newcommand\theorygcin{\m{\theory_{gcin}}}
\newcommand\theorystar{\m{\theory^*}}
\newcommand\theorygcstar{\m{\theory_{gc}^*}}

\newcommand\structin{\m{\struct_{in}}}
\newcommand\structout{\m{\struct_{out}}}
\newcommand\structgc{\m{\struct_{gc}}}
\newcommand\structgcin{\m{\struct_{gcin}}}
\newcommand\structgcout{\m{\struct_{gcout}}}
\newcommand\structinstar{\m{\struct_{in}^*}}
\newcommand\structgcinstar{\m{\struct_{gcin}^*}}

\newcommand\sym{\m{\sigma}}
\newcommand\globdomsym[1]{\m{\sigma_{#1}}}
\newcommand\locdomsym[2]{\m{\sigma_{#1}^{#2}}}
\newcommand\locdomintch[2]{\m{\mathbb{G}_{#1}^{#2}}}
\newcommand\dpg[2]{\m{DPG(#1,#2)}}
\newcommand\argnode[2]{\m{#1.#2}}

\newcommand\successor[1]{\m{#1'}}
\newcommand\erasethisproof[2]{#2}

\newcommand\kodkod{\logicname{Kodkod}}
\newcommand\clasp{\logicname{clasp}}
\newcommand\clingo{\logicname{clingo}}
\newcommand\SEM{\logicname{SEM}}
\newcommand\idpsym{\logicname{IDPsym}}

\newcommand\graceful{\textbf{graceful}}
\newcommand\holes{\textbf{pigeons}}
\newcommand\nqueens{\textbf{200queens}}
\newcommand\crew{\textbf{crew}}

\newtheorem{theorem}{Theorem}[section]
\newtheorem{lemma}[theorem]{Lemma}
\newtheorem{proposition}[theorem]{Proposition}
\newtheorem{corollary}[theorem]{Corollary}

\newtheorem{definition}[theorem]{Definition}
\newtheorem{example}[theorem]{Example}
\newtheorem{problem}[theorem]{Problem}
\newtheorem{remark}[theorem]{Remark}


\newcommand\xqed[1]{
  \leavevmode\unskip\penalty9999 \hbox{}\nobreak\hfill
  \quad\hbox{#1}}
\newcommand\demo{\xqed{$\triangledown$}}

\newcommand{\arxiv}[3]{
	\ifthenelse{\boolean{nocomments}}
	{} 
	{ 
		\ifthenelse{\boolean{commentmargin}}
			{ {\color{#3} \marginpar{\color{#3}\sc #2}#1}  } 
			{  {\color{#3} {\sc #2}: #1}  } 
	}
}

\setboolean{nocomments}{true}


\begin{abstract}
Symmetry in combinatorial problems is an extensively studied topic. We continue this research in the context of model expansion problems, with the aim of automating the workflow of detecting and breaking symmetry. We focus on \emph{local domain symmetry}, which is induced by permutations of domain elements, and which can be detected on a first-order level. As such, our work is a continuation of the symmetry exploitation techniques of model generation systems, while it differs from more recent symmetry breaking techniques in answer set programming which detect symmetry on ground programs. Our main contributions are sufficient conditions for symmetry of model expansion problems, the identification of \emph{local domain interchangeability}, which can often be broken completely, and efficient symmetry detection algorithms for both local domain interchangeability as well as local domain symmetry in general. Our approach is implemented in the model expansion system \idp, and we present experimental results showcasing the strong and weak points of our approach compared to \sbass, a symmetry breaking technique for answer set programming.

\textbf{Under consideration for acceptance in TPLP.}

\end{abstract}

\section{Introduction}
\label{sec:intro}
Many problems exhibit symmetry. For instance, the set of trucks in a routing problem is interchangeable, a chess board can be mirrored onto itself, an input graph has non-trivial automorphisms, etc.
It is a well-known burden of combinatorial search engines that they visit each of the (potentially exponentially many) symmetric areas of their search space, and hence waste valuable time rediscovering already known information. 
In order to solve this problem, research on symmetry has been extensive, especially in the constraint programming (CP)~\cite{gent2006symmetry} and satisfiability solving (SAT) community~\cite{Sak09HBSAT}.

For logic-based systems, much work has been done in the context of theorem proving and finite model generation systems~\cite{Zhang95sem,Audemard02reasoningby,Claessen03newtechniques,tacas/TorlakJ07}. With the advent of answer set programming (ASP), interest in symmetry for logics is renewed~\cite{drtiwa11a}.

In this paper, we continue research on symmetry in classical logic, with a focus on symmetry for model expansion problems. 
We propose the notion of \emph{local domain symmetry} in Section~\ref{sec:symmetries}, a common form of symmetry stemming from permutations of domain elements.
We show in Section~\ref{sec:breaking} how \emph{local domain interchangeability}, a particular type of local domain symmetry, can be broken completely with a linear number of symmetry breaking constraints.
Section~\ref{sec:detection} gives a detection algorithm for local domain symmetry in general, and for local domain interchangeability in particular.
These detection algorithms operate on a first-order level, hence they avoid the computational blow-up of a \emph{ground} theory.
In Section~6, we experimentally compare our approach to state-of-the-art ASP symmetry
breaking which performs symmetry detection on the ground theory. These
experiments confirm that symmetry detection at the first order level is
indeed faster, and in some cases, we achieve stronger symmetry breaking.
However, there are also cases where less symmetry is detected.
We conclude in Section~\ref{sec:conclusion}.
Proofs are postponed to \ref{app:proofs}.


\section{Preliminaries}
\label{sec:prelims}
We assume familiarity with the basics of first-order logic.
A \emph{vocabulary} \voc is a set of \emph{predicate symbols} $P/n$ of arity $n \geq 0$ and \emph{function symbols} $f/n$ of arity $n\geq 0$.
Often, we will simply refer to a \emph{symbol} $S/n \in \voc$, which represents an $n$-ary predicate or function symbol.
\emph{Variables}, \emph{terms}, \emph{atoms}, \emph{quantifiers}, \emph{formulas} and \emph{theories} are defined as usual \cite{Enderton01}.
A $\voc$-theory \theory is a theory with vocabulary $\voc$, i.e., such that the free symbols of $\theory$ are all in $\voc$.

Slightly deviating from the standard presentation of first-order logic, we consider both variables and constants to be function symbols of arity $0$, as they both serve to identify a single domain element.
Without loss of generality, we assume variables are \emph{renamed apart}, i.e., each variable is bound by at most one quantifier.
This simplifies the presentation of our results.

A \emph{\voc-structure} \struct consists of a domain \domain and an \emph{interpretation} to each symbol in $\voc$.
For each $n$-ary predicate symbol $P$, interpretation $P^\struct$ is an $n$-ary relation on $\domain$, i.e., $P^\struct \subseteq \domain^n$. 
For an $n$-ary function $f$, interpretation $f^\struct$ is an $n$-ary function $\domain^n\to\domain$.
If \voc-structure \struct and $\voc'$-structure $\struct'$ have the same domain \domain and $\voc \cap \voc'=\emptyset$, $\struct \sqcup \struct'$ is the $\voc \cup \voc'$-structure over \domain that interprets all symbols in $\voc$ according to $\struct$ and all symbols in $\voc'$ according to $\struct'$.
If $\struct$ is a \voc-structure with domain \domain, $x$ a $0$-ary function symbol, and $d\in \domain$, then we use $\struct[x:d]$ for the $\voc\cup\{x\}$-structure that equals $\struct$, except for interpreting $x$ by $d$.
Because variables are considered $0$-ary function symbols as well, the notion of ``variable assignment'', which is used in the classical presentation of first-order logic, is subsumed by our notion of structure extension $\struct[x:d]$.

The \emph{value} $t^\struct$ of a term $t$ in a structure \struct is a domain element $d \in \domain$, and is defined as usual (assuming \struct interprets all  symbols in $t$).
The \emph{truth value} $\f^\struct$ of a formula \f in a structure \struct is either true or false, and is also defined as usual.
A \voc-structure \struct is a \emph{model} of a \voc-theory \theory
(written as $\struct \models \theory$) if for each formula \f in
\theory, $\f^\struct$ is true.
Many combinatorial problems can be conveniently modelled as a \emph{model expansion} problem $MX(\theory,\structin)$, where \theory is a \voc-theory \theory and $\structin$ is a $\vocin$-structure with $\vocin\subseteq \voc$.
We refer to $\vocin$ as the \emph{input vocabulary}, and $\vocout=\voc \setminus \vocin$ as the \emph{output vocabulary}.
A \emph{solution} to a model expansion problem $MX(\theory,\structin)$ with output vocabulary \vocout is a \vocout-structure \structout (sharing \structin's domain) such that $\structin \sqcup \structout \models \theory$; $\structin \sqcup \structout$ is a model to \theory that \emph{expands} \structin.



\section{Symmetries}
\label{sec:symmetries}
Throughout this section, we assume a fixed domain \domain and use
$\strucs$ to refer to the set of all structures with domain \domain.

As a running example, we use a simple graph coloring problem. 
\begin{example}
\label{ex:graph_coloring_start}
Let $\voc_{gc}$ be the vocabulary consisting of predicate symbols $V/1$, $C/1$, $Edge/2$ and a function symbol $Color/1$. 
A valid colored graph is expressed by the theory $\theorygc$:
\begin{ltheo}
	\forall x_1 ~ y_1\colon Edge(x_1,y_1) \limplies (Color(x_1) \neq Color(y_1)) \\
	\forall x_2 ~ y_2\colon Edge(x_2,y_2) \limplies V(x_2) \land V(y_2) \\
	\forall x_3\colon C(Color(x_3))
\end{ltheo}
Let $\voc_{gcin}=\voc_{gc}\setminus\{Color/1\}$. 
Input data containing vertices, colors and a graph is expressed as a $\vocgcin$-structure $\structgcin$ with domain $\domain=\{t,u,v,w,r,g,b\}$, with interpretations
\begin{ltheo}
V^{\structgcin}=\{t,u,v,w\}~~
Edge^{\structgcin}=\{(t,u),(u,v),(v,w),(w,t)\}~~
C^{\structgcin}=\{r,g,b\}.
\end{ltheo}

The model expansion problem $MX(\theorygc,\structgcin)$ now consists of finding a $\vocgcout=\{Color/1\}$-structure $\structgcout$ such that $\structgcin \sqcup \structgcout \models \theorygc$.
We let $\structgcout$ contain the interpretation
\[ Color^{\structgcout}=t\mapsto r,u\mapsto g,v \mapsto b, w\mapsto g, r\mapsto r, g\mapsto g, b \mapsto b \]
which represents a valid coloring of the input graph.
Indeed, $\structgc=\structgcin \sqcup \structgcout\models \theorygc$.
\demo\end{example}

\subsection{Symmetry of a theory}
\label{sec:theorysymmetry}
\begin{definition}[Symmetry]
  \label{newsymmetry_definition}
  A mapping $\sigma \colon \strucs \to \strucs$ is a {\em structure
    transformation.} A structure transformation $\sigma$ is a {\em
    symmetry} for \voc-theory \theory if for all \voc-structures $\struct \in
  \strucs$, $\struct\models\theory$ iff
  $\sigma(\struct)\models\theory$.
\end{definition}

The set of symmetries of \theory forms a group under composition ($\circ$).
In this paper, we study how to detect and exploit symmetries. 
Detecting all symmetries of a theory is computationally at least as hard as deciding whether the theory is satisfiable (if not, all structure transformations are symmetries). 
Instead, we focus on symmetries that can be detected by means of syntactical analysis, and that are induced by permutations of domain elements.

\begin{definition}[Domain permutation]
  A bijection $\pi \colon \domain \to \domain$ is a \emph{domain
  permutation}. 
  A domain permutation \emph{induces a structure transformation} $\locdomsym{\pi}{}$: for each predicate symbol $P/n$, $(\pi(d_1), \ldots ,\pi(d_n))
  \in P^{\locdomsym{\pi}{}(\struct)}$ iff $(d_1, \ldots ,d_n) \in
  P^\struct$, and for each function symbol $f/n$, $f^{\sigma_{\pi}(\struct)}(\pi(d_1), \ldots
  ,\pi(d_n)) = \pi(d) $ iff $f^\struct(d_1, \ldots
  ,d_n) = d  $.
\end{definition}


\begin{proposition} 
  Any structure transformation induced by a domain permutation is a symmetry for any theory.
\end{proposition}

We call this type of symmetry induced by only a domain permutation {\em global domain symmetry}.

We use cycle notation to compactly represent permutations, e.g., $(a~b~c)(d~e)$ is a permutation that maps element $a$ to $b$, $b$ to $c$, $c$ to $a$, swaps $d$ and $e$, and maps any other element to itself.

\begin{example}[Example~\ref{ex:graph_coloring_start} continued]
\label{ex:newgraph_coloring_2}
The domain permutation $(v~r)$ induces a global domain symmetry
$\globdomsym{(v~r)}$ of \theorygc. 
$\globdomsym{(v~r)}(\structgc)$ gives
\begin{ltheo}
\domain=\{t,u,v,w,r,g,b\} ~~~~~
V^{\globdomsym{(v~r)}(\structgc)}=\{t,u,r,w\} ~~~~~
Edge^{\globdomsym{(v~r)}(\structgc)}=\{(t,u),(u,r),(r,w),(w,t)\} \\
C^{\globdomsym{(v~r)}(\structgc)}=\{v,g,b\} ~~~~~~
Color^{\globdomsym{(v~r)}(\structgc)}=t\mapsto v,u\mapsto g,r\mapsto b,w\mapsto g,v \mapsto v, g \mapsto g, b \mapsto b,
\end{ltheo}
which is still a model of $\theorygc$ (though $r$ now acts as a vertex and $v$ as a color).
\demo\end{example}

Finite model generators such as
  \kodkod~\cite{tacas/TorlakJ07} or \SEM~\cite{Zhang95sem} focus on the task of
  generating a model with a given domain for a given theory.
Since every domain permutation induces a global domain symmetry, these systems 
have mechanisms to cope with global domain symmetry.

However, a global domain symmetry $\locdomsym{\pi}{}$ is a rather restrictive
concept as it applies $\pi$ on every argument of every tuple in
every interpretation of a structure. A larger class of transformations can be
described when $\pi$ is only applied locally. For example, one
could apply $\pi$ only on the interpretation of some symbols, or
even more 
fine-grained, 
only on some of the arguments in the tuples of an interpretation. Given a predicate or function symbol $S/n$, we use $\argpos{S}{i}$ with $1\leq  i \leq n$ to denote the $i^{th}$ \emph{argument position} of $S$; if $S$ is a function symbol, we use $S|0$ for the output argument of $S$. Note that variables, being treated as function symbols, also form argument positions.

\begin{definition}[Structure transformation induced by $A,\pi$]
\label{def:inducedstructuretransformation}
  Let $\pi$ be a domain permutation and $A$ a set of argument
  positions. 
  The structure transformation $\locdomsym{\pi}{A}$ \emph{induced by $A,\pi$} is defined by
  \begin{align*}
  (\localperm{}{P}{1}(d_1),\ldots,\localperm{}{P}{n}(d_n))\in P^{\locdomsym{\pi}{A}(\struct)}&\text{ iff } (d_1,\ldots, d_n)\in P^\struct\\
  f^{\locdomsym{\pi}{A}(\struct)}(\localperm{}{f}{1}(d_1), \ldots
  ,\localperm{}{f}{n}(d_n)) = \localperm{}{f}{0}(d_0) &\text{ iff } f^\struct(d_1, \ldots
  ,d_n) = d_0
  \end{align*}
  where $\localperm{}{S}{i}(d) = \pi(d)$ if $\argpos{S}{i} \in A$ and $\localperm{}{S}{i}(d) =d$ otherwise.
\end{definition}
Thus, for each domain tuple in the interpretation of a symbol $S$, the structure transformation induced by $A,\pi$ only applies $\pi$ to domain elements that occur at an index $i$ corresponding to an argument position $\argpos{S}{i} \in A$.
Note that if $A$ contains argument positions over symbols $S$ not interpreted by $\struct$ (e.g., variable symbols), those argument positions are simply ignored by $\locdomsym{\pi}{A}$.

\begin{definition}[Local domain symmetry]
  Let $\theory$ be a theory. 
  A \emph{local domain symmetry} for $\theory$ is a structure transformation induced by a set of argument positions $A$ and a domain permutation $\pi$, that also is a symmetry for $\theory$.
\end{definition}


A global domain symmetry $\globdomsym{\pi}$ for a $\voc$-theory is a local domain symmetry $\locdomsym{\pi}{A}$ where $A$ includes all argument positions of all symbols in $\voc$.
As such, local domain symmetry is a generalization of global domain symmetry, and allows us to detect and exploit more symmetry.
However, not all $A,\pi$-induced structure transformations are symmetries. 
Below, we propose a syntactic criterion to identify a set of argument positions $A$ that guarantees that $\locdomsym{\pi}{A}$ is a symmetry for a given theory. 
Intuitively, the criterion can be formulated as follows: whenever a term $f(\ldots)$ occurs as the $i$'th argument in a predicate or function symbol $S$, then $\argpos{f}{0} \in A$ if and only if $\argpos{S}{i} \in A$. 

\begin{definition}
 Let $\theory$ be a theory.
 Assume $\argpos{f}{0}$ and $\argpos{S}{i}$ are argument positions with $S$ either a predicate or a function symbol. We call $\argpos{f}{0}$ and $\argpos{S}{i}$ \emph{directly connected by \theory} if one of the following holds:
 \begin{compactitem}
  \item an expression $S(t_1,\dots, t_{i-1},f(\ttt'),t_{i+1},\dots,t_n)$ occurs in \theory, or
  \item $i=0$ and an expression $S(\ttt)= f(\ttt')$ occurs in \theory. 
 \end{compactitem}
 A set $A$ of argument positions is \emph{connectively closed under \theory} if for each $\argpos{S}{i} \in A$, each argument position $\argpos{R}{j}$ directly connected to $\argpos{S}{i}$ by \theory, is also in $A$. 
\end{definition}

\begin{example}[Example~\ref{ex:graph_coloring_start} continued]
\label{ex:graph_coloring_3}
According to the first formula in $\theorygc$, 
$\argpos{x_1}{0}$ is directly connected to $\argpos{Edge}{1}$ and $\argpos{Color}{1}$, 
while $\argpos{y_1}{0}$ is directly connected to $\argpos{Edge}{2}$ and $\argpos{Color}{1}$. 
Analyzing all formulas, we
find the following two sets are connectively closed under $\theorygc$:
$A=\{\argpos{C}{1},\argpos{Color}{0}\}$ and
$B=\{\argpos{V}{1},$ $\argpos{Edge}{1},$ $\argpos{Edge}{2},$ $\argpos{Color}{1},$ $ \argpos{x_1}{0},$ $\argpos{y_1}{0},$ $ \argpos{x_2}{0},$ $\argpos{y_2}{0},$ $\argpos{x_3}{0}\}$.

Applying the induced structure transformation $\locdomsym{(v~r)}{A}$ on $\structgc$ gives
\begin{ltheo}
\domain=\{t,u,v,w,r,g,b\} ~~~
V^{\locdomsym{(v~r)}{A}(\structgc)}=\{t,u,v,w\} ~~~
Edge^{\locdomsym{(v~r)}{A}(\structgc)}=\{(t,u),(u,v),(v,w),(w,t)\} \\
C^{\locdomsym{(v~r)}{A}(\structgc)}=\{v,g,b\} ~~~
Color^{\locdomsym{(v~r)}{A}(\structgc)}=t\mapsto v,u\mapsto g,v\mapsto b,w\mapsto g,r \mapsto v, g \mapsto g, b \mapsto b
\end{ltheo}
which is also a model of $\theorygc$ (here, domain element $v$ serves both as a
vertex and a color).
\demo\end{example}


\begin{theorem}[Local domain symmetry condition]
\label{thm:locdomsymcondition}
Let \voc be a vocabulary, \theory a theory over $\voc$, $\pi$ a domain permutation and $A$ a set of argument positions. If $A$ is connectively closed under \theory, then $\locdomsym{\pi}{A}$ is a local
domain symmetry of \theory.
\end{theorem}

This theorem is useful when detecting symmetry for model expansion problems with an empty input structure, but it will also prove useful for non-empty input structures.

\begin{example}[Example~\ref{ex:graph_coloring_3} continued]
\label{ex:graph_coloring_4}
The argument position set $A$
induces local domain symmetries that correspond to permuting the
colors of a graph coloring problem, while
$B$
induces symmetry on the vertices and $A \cup B$ induces global domain symmetries.
%
\demo\end{example}

\subsection{Symmetry for model expansion}
\label{sec:inputsymmetry}
Recall that a model expansion problem $MX(\theory,\structin)$ consists of finding structures $\structout$ such that $\structin \sqcup \structout \models \theory$.

\begin{definition}[Symmetry for MX]
\label{def:symformx}
Let $MX(\theory,\structin)$ be a model expansion problem with output vocabulary \vocout, and let $\Gamma_\domain^{\vocout}$ be the set of \vocout-structures with domain $\domain$.
A structure transformation $\sigma \colon \Gamma_\domain^{\vocout} \to \Gamma_\domain^{\vocout}$ is a symmetry of $MX(\theory,\structin)$ if for each $\structout \in \Gamma_\domain^{\vocout}$, $\structin \sqcup \structout \models \theory$ iff $\structin \sqcup \sigma(\structout) \models \theory$.
\end{definition}

Analogous to Definition~\ref{def:inducedstructuretransformation}, a domain permutation $\pi$ and argument position set $A$ induce a structure transformation $\locdomsym{\pi}{A}$ on $\Gamma_{\domain}^{\vocout}$. We call $\locdomsym{\pi}{A}$ a \emph{local domain symmetry of $MX(\theory,\structin)$} if $\locdomsym{\pi}{A}$ is a symmetry of $MX(\theory,\structin)$.


\begin{example}[Example~\ref{ex:graph_coloring_start} continued]
\label{ex:graph_coloring_5}
Let $A$ be the argument position set $\{\argpos{V}{1},\argpos{Edge}{1},\argpos{Edge}{2},\argpos{Color}{1},\argpos{x_1}{0},\argpos{y_1}{0},\argpos{x_2}{0},\argpos{y_2}{0},\argpos{x_3}{0}\}$. 
Observe that $A$ is connectively closed under \theorygc and that the induced structure transformation $\locdomsym{(t~u~v~w)}{A}$ is a local domain symmetry of $MX(\theorygc,\structgcin)$.
However, connectively closedness under the theory is neither a sufficient nor a necessary condition for an $A,\pi$-induced structure transformation to be a symmetry of a model expansion problem.

For instance, argument position set $B=\{\argpos{Edge}{1},\argpos{Edge}{2},\argpos{Color}{1},\argpos{x_1}{0},\argpos{y_1}{0},\argpos{x_3}{0}\}$ is not connectively closed under $\theorygc$, though $\locdomsym{(t~u~v~w)}{B}$ is still a symmetry of $MX(\theorygc,\structgcin)$.

Moreover, since $A$ is connectively closed, $\locdomsym{(v~r)}{A}$ is a local domain symmetry of $\theorygc$, but it is not a symmetry of $MX(\theorygc,\structgcin)$.
Indeed, 
\[ Color^{\locdomsym{(v~r)}{A}(\structgcout)}=t\mapsto v,u\mapsto g,v\mapsto b,w\mapsto g,r\mapsto v, g \mapsto g, b \mapsto b \]
maps $t$ and $r$ to node $v$, which is not consistent with $\forall x_3 \colon C(Color(x_3))$ and $C^\structgcin$.
\demo\end{example}

The above example shows that for model expansion, local domain symmetries are useful, but Theorem~\ref{thm:locdomsymcondition} does not suffice to identify them.
Below, we give a sufficient condition for $A,\pi$-induced structure transformations to be local domain symmetries of a model expansion problem.
For this, we require the notion of a \emph{decomposition}. 

\begin{definition}
Let $MX(\theory,\structin)$ be a model expansion problem with input vocabulary \vocin.
Also, let $\theorystar$ be equal to $\theory$ with each occurrence of a symbol $S \in \vocin$ replaced by a unique new copy $S_i$, let $\vocinstar$ be the vocabulary containing all copy symbols $S_i$, and let $\structinstar$ be the $\vocinstar$-structure where for each copy $S_i$, $S_i^{\structinstar}=S^\structin$.
We call $MX(\theorystar,\structinstar)$ the \emph{decomposition} of $MX(\theory,\structin)$.
\end{definition}
It is clear that a model expansion problem and its decomposition have the same solutions, as they have the same output vocabulary and as each occurrence of a copy $S_i$ in $\theorystar$ imposes the same constraints on models for $\theorystar$ as $S$ did for $\theory$ (since $S_i^{\structinstar}=S^\structin$).

\begin{example}[Example~\ref{ex:graph_coloring_5} continued]
\label{ex:graph_coloring_6}
Let $MX(\theorygcstar,\structgcinstar)$ be the decomposition of $MX(\theorygc,\structgcin)$. \theorygcstar consists of
 \begin{ltheo}
 	\forall x_1 ~ y_1\colon Edge_1(x_1,y_1) \limplies (Color(x_1) \neq Color(y_1)) \\
 	\forall x_2 ~ y_2\colon Edge_2(x_2,y_2) \limplies V_1(x_2) \land V_2(y_2) \\
 	\forall x_3\colon C_1(Color(x_3))
 \end{ltheo}
\demo\end{example}
\begin{theorem}[Local domain symmetry condition for $MX$]
\label{thm:locdomsymconditionmx}
Let $MX(\theory,\structin)$ be a model expansion problem with decomposition $MX(\theorystar,\structinstar)$.
If $A$ is connectively closed under $\theorystar$ and $\locdomsym{\pi}{A}(\structinstar)=\structinstar$
then $\locdomsym{\pi}{A}$ is a symmetry for $MX(\theory,\structin)$.
\end{theorem}

\begin{example}[Example~\ref{ex:graph_coloring_6} continued]
\label{ex:graph_coloring_7}
Argument position set $A=\{\argpos{Edge_1}{1},\argpos{Edge_1}{2},\argpos{Color}{1},\argpos{x_1}{0},\argpos{y_1}{0},\argpos{x_3}{0}\}$ is connectively closed under $\theorygcstar$, and $\locdomsym{(t~u~v~w)}{A}(\structgcinstar)=\structgcinstar$.
Thus, $\locdomsym{(t~u~v~w)}{A}$ is a symmetry of $MX(\theorygc,\structgcin)$, exploiting cyclicity of the input graph.
However, $\locdomsym{(t~u)}{A}$ is not a symmetry of $MX(\theorygc,\structgcin)$, as the input interpretation of $Edge_1$ is not preserved by swapping $t$ and $u$.
\demo\end{example}

Note that if $\vocin=\emptyset$, the conditions of Theorem~\ref{thm:locdomsymconditionmx} degenerate into the conditions of Theorem~\ref{thm:locdomsymcondition}. 
Also, for $\locdomsym{\pi}{A}$ satisfying Theorem~\ref{thm:locdomsymconditionmx}, $A$ typically contains argument positions over both $\vocout$ and the decomposed $\vocinstar$ (as well as over variables).
%
Lastly, the requirement that $A$ is connectively closed under $\theorystar$ is weaker than being closed under $\theory$.
For example, let $\theory$ be $ P(f) \lor P(g)$, with only $P$ interpreted by the input structure. 
The only connectively closed set under $\theory$ is $\{\argpos{P}{1},\argpos{f}{0},\argpos{g}{0}\}$.
However, under the corresponding decomposition theory $\theorystar= P_1(f) \lor P_2(g)$, there are three connectively closed sets:
$\{\argpos{P_1}{1},\argpos{f}{0}\}$, $\{\argpos{P_2}{1},\argpos{g}{0}\}$ and their union.

\subsection{Subdomain interchangeability}
Local domain symmetries for a theory \theory can be
identified by computing argument position sets $A$ that are connectively closed. 
Then, as mentioned in Section~\ref{sec:theorysymmetry}, any
permutation $\pi$ of the domain $\domain$ gives rise to a local domain
symmetry $\locdomsym{\pi}{A}$. 
For model expansion, $\locdomsym{\pi}{A}$ must preserve the input structure, so not all $\pi$ are guaranteed to induce symmetry.
However, given a suitable set of argument positions $A$, a subdomain $\delta \subseteq \domain$ might exist for which any permutation of $\delta$ induces a symmetry of the model expansion problem.

\begin{definition}[$A$-interchangeable subdomain]
\label{def:locdom_interchangeability}
Let $MX(\theory,\structin)$ be a model expansion problem, $A$ a
set of argument positions
and $\delta$ a subset of the domain. 
$\delta$ is an {\em $A$-interchangeable subdomain} if for every permutation $\pi$ over $\delta$, the structure transformation $\locdomsym{\pi}{A}$ induced by $A,\pi$ is a local domain symmetry for $MX(\theory,\structin)$.
The \emph{subdomain interchangeability group} $\locdomintch{\delta}{A}$ is the group of all local domain symmetries induced by an $A$-interchangeable subdomain $\delta$.
\end{definition}

\begin{example}[Example~\ref{ex:graph_coloring_start} continued]
\label{ex:graph_coloring_8}
Given $MX(\theorygc,\structgcin)$, $\{r,g,b\}$ and $\{t,u,v,w\}$ are $A$-interchangeable subdomains for $A=\{\argpos{C}{1},$ $\argpos{Color}{0}\}$.
For $B=\{\argpos{V}{1},\argpos{Edge}{1},\argpos{Edge}{2},\argpos{Color}{1}, \argpos{x_1}{0},\argpos{y_1}{0}, \argpos{x_2}{0},\argpos{y_2}{0},\argpos{x_3}{0}\}$, $\{r,g,b\}$ is a $B$-in\-ter\-changeable subdomain. However, as $\locdomsym{(t~u)}{B}$ does not preserve the interpretation of $Edge$, $\{t,u,v,$ $w\}$ is not a $B$-interchangeable subdomain.
\demo\end{example}

Many problems, when modelled as a model expansion problem, exhibit subdomain interchangeability. For instance, a set of nurses in a scheduling problem, a set of colors in a graph coloring problem, or a set of trucks in a planning problem often are interchangeable subdomains.

Subdomain interchangeability groups contain a number of symmetries factorial in the size of the interchangeable subdomain, leading to an exponential slowdown of many combinatorial search algorithms.
However, as we show in Section~\ref{sec:breaking}, many subdomain interchangeability groups can be completely broken with a number of constraints linear in the size of the subdomain.

\subsection{More symmetry}
\label{sec:more_symmetry}
Even though local domain symmetry is a useful form of symmetry, it does not capture all symmetry properties that might be present in a model expansion problem.

\begin{example}[Example~\ref{ex:graph_coloring_start} continued]
\label{ex:graph_coloring_9}
The graph coloring problem $MX(\theorygc,\structgcin)$ asks to color a circular directed graph of $4$ vertices $\{t,u,v,w\}$.
Note that given any satisfying coloring for this graph, swapping the colors of $t$ and $v$ (or $u$ and $w$) keeps $\theorygc$ satisfied.
This is a clear symmetry property of the graph coloring instance, but it cannot be captured using the notion of local domain symmetry as defined in this paper.
For instance, if we take the argument position set $A=\{\argpos{V}{1},\argpos{Edge}{1},\argpos{Edge}{2},\argpos{Color}{1}\}$ representing symmetry in the vertices, then the induced structure transformation $\locdomsym{(t~v)}{A}$ is not a symmetry of $MX(\theorygc,\structgcin)$ since it does not preserve the interpretation of $Edge$.

One way to fix this is 
%
based on the observation that argument positions $\argpos{Edge}{1}$ and $\argpos{Edge}{2}$ are indistinguishable in $\theorygc$: in each sentence of $\theorygc$, one could swap any quantifier over $\argpos{Edge}{1}$ with one over $\argpos{Edge}{2}$, ending up with a sentence equivalent to the original one.
In more bold words, argument positions $\argpos{Edge}{1}$ and $\argpos{Edge}{2}$ \emph{themselves} are symmetric.
This symmetry property can be captured by generalizing the notion of an $A,\pi$-induced structure transformation to allow for swaps or permutations of argument positions.

For instance, we could define $\locdomsym{A}{(\argpos{Edge}{1}~\argpos{Edge}{2})(t~v)}$ to first map $Edge^{\structgcin}$ to $\{(e,d)~|~(d,e) \in Edge^{\structgcin}\}$ before applying $\locdomsym{A}{(t~v)}$.
Note that $\locdomsym{A}{(\argpos{Edge}{1}~\argpos{Edge}{2})(t~v)}$ would preserve $Edge^{\structgcin}$ and $\structgcin$ in general, while also preserving satisfaction to $\theorygc$, making it a symmetry of $MX(\theorygc,\structgc)$.
\demo\end{example}

Similarly, problems with spatial properties often have rotational or reflectional symmetry, which is not covered by the presented notion of local domain symmetry. One such example is the N-Queens problem, which is experimentally investigated in Section \ref{sec:experiments}.

\ignore{
\subsection{Relation to typed logic}
\jo{also mention stability of symmetry under transformations}

The local domain symmetry condition presented in this section has a very strong relation to types, as permuting the domain elements in a structure only on certain argument positions is very similar to only permuting a certain type domain in typed logic. 
This is no coincidence, as types often represent distinct sets of symmetric objects from a problem domain. 
\cite{Claessen03newtechniques} showed that automatic type inference allows for more powerful symmetry exploitation in model generation problems. Our work extends this to the context of model expansion, where an input structure influences the connectedness of argument positions, and hence types.

\subsubsection{Types}
\maurice{naar hier verhuisd}

\jo{Mention types: yes, we are essentially deriving a maximally fine-grained type system. This has been done before. However, we continue in this formalism because types should represent sets of real-life objects (KR argument), users should not be trusted in optimally typing their theory (usability argument), even optimal types can get split given an interpreted symbol (see example in next section) (MX argument) (imagine we add a predicate $Independent(Vertex)$ to the graph coloring problem. Same type, but different symmetry group than $Color/1$ if $Edge/2$ is interpreted).}

\subsection{Symmetry and typed logic}
\label{subsection_symmetry_and_types}
Local domain symmetry has an interesting correspondence with typed logic. Typed logic introduces \jo{a finite set of} domains $\domain_1,\domain_2,\ldots$ -- called \emph{types} -- instead of only one domain $\domain$. Each structure $\struct$ provides an actual set of domain elements $\domain_i^\struct$ for each type. Each argument position of a predicate or function symbol is associated with a type. The interpretation of a predicate $P$ in \struct would then be $P^\struct \subseteq \domain_1^\struct \times \ldots \times \domain_n^\struct$, where $\domain_i$ is the type associated with $(P,i)$. \jo{ref naar typed logic?}

Well-formed theories over typed logic satisfy the following property:
\begin{compactitem}
\item for each occurrence in \theory of a term $t$ on argument position $\argpos{S}{i}$, it holds that $(\symb{t},0)$ and $\argpos{S}{i}$ have the same type.
\item for each atom $t_1=t_2$ in \theory, it holds that $(\symb{t_1},0)$ and $(\symb{t_2},0)$ have the same type.
\end{compactitem}
The similarity with the local symmetry condition is no coincidence: one of the most common causes of symmetry is a set of mutually interchangeable objects over which relations have to be constructed, but the constraints never distinguish between individual objects. A classical example is the set of colors in the graph coloring problem, the set of pigeons or holes in the pigeonhole problem, a fleet of identical vehicles in a vehicle routing problem, etc.

When modelling these problems in typed logic, such sets of objects manifest themselves as interpretations to types. Hence, in the case of typed logic, it is natural to limit domain permutations $\pi$ to only the interpretation of a single type, prohibiting domain permutations that permute domain elements of different types simultaneously. This coincides with the notion that a symmetry should respect the interpretations of a given partial structure, which in the typed case includes interpretations of types.

\jo{We could filosofize even more, stating that any local domain symmetry that respects the given type interpretations but permutes domain elements from different types is a composition of local domain symmetries that permute only one domain. Maybe there is even more. Don't think it is interesting enough to mention.}

\subsubsection{Logical rules}

\maurice{naar hier verhuisd}

One intuition on why Theorem~\ref{thm:locdomsymcondition} holds is that the semantics of first-order logic treat all domain elements uniformly identically.
Hence, renaming the domain does not fundamentally alter the set of models for a theory.
The connectedness property of the set of argument positions then functions to make sure any cardinality constraints remain satisfied.

The semantics of logical rules have the same property: the (three-valued) immediate consequence operator used in stable semantics or well-founded semantics \jo{cite} makes no distinction between domain elements.
Hence, local domain symmetry properties are relevant in logical theories with stable or well-founded semantics as well.

\subsection{Extensions of FO}
For many practical applications, pure first-order logic does not suffice as a modelling language. Therefore, people have integrated FO with various language constructs such as rule sets (under stable or well-founded semantics)~\cite{marek99stable,\refto{fodot}}, aggregates~\cite{phd/Marien09}, arithmetic~\mycite{CPsupport} etc. 
The local domain symmetry condition presented in this paper can easily be extended to those settings. For instance, incorporating arithmetic simply involves assuring that pre-interpreted inequality symbols or arithmetic functions are left fixed by local domain symmetries.
As for logic rules, the stable model or well-founded semantics is typically defined on propositional rules, which are derived by filling in domain elements in quantifiers, which introduces no new asymmetry in the rules. Hence, local domain symmetry is not limited to classical FO but is also useful for these semantics. 

As such, our criteria for detecting symmetry are also applicable to answer set programming \mycite{ASP} or in constraint programming \mycite{CP}.


Thirdly, \textbf{inductive definitions} consist of sets of logical rules of the form
\[ \forall \bar{x}\colon P(\bar{x}) \leftarrow \f(\bar{x}) \]
which follow the \emph{well-founded} semantics \jo{ref}. The well-founded semantics is closely related to stable-model semantics as employed by ASP systems or the semantics of Prolog programs. 
As far as symmetry properties is concerned, inductive definitions are constructed from atoms, for which the well-founded semantics and classical semantics coincide, so that Lemma~\ref{lemma_symmetry_of_atom} also holds for atoms occurring in inductive definitions. \jo{todo: argument why inductive definitions are symmetrical for local domain symmetry.}
}

\section{Symmetry breaking and local domain interchangeability}
\label{sec:breaking}

A standard approach of dealing with symmetry extends the theory with
\emph{symmetry breaking constraints} that eliminate symmetric
solutions while guaranteeing that at least one solution to the
original problem is preserved (if it exists). This way, a search
algorithm will not get stuck in parts of the search space symmetric to those already explored. 
A set of symmetry breaking constraints $\varphi$ is {\em sound} for a symmetry group $\mathbb{G}$ if for each solution $\struct$, there exists \emph{at least} one $\sigma \in \mathbb{G}$ such that $\sigma(\struct)$ satisfies $\varphi$; it is \emph{complete}
if there exists \emph{at most} one such $\sigma \in \mathbb{G}$~\cite{walsh_recent_results_2012}.

Often, symmetry breaking is done by defining a lexicographical order over the set of 
candidate solutions. For a given symmetry, so-called \emph{lex-leader constraints} then encode that each solution's symmetrical image cannot be strictly smaller under the defined lexicographical order.
As long as the chosen lexicographical order is fixed, the conjunction of lex-leader constraints for any set of symmetries is sound.

In a model expansion context, the set of candidate solutions $\Gamma_\domain^\vocout$ consists of all $\vocout$-structures with domain \domain. A logical formula that is added to the theory takes over the role of a constraint.
We construct a lexicographical order $\preceq_\Gamma$ over $\Gamma_\domain^\vocout$ from an order $\preceq_\domain$ over \domain and an order $\preceq_{\vocout}$ over \vocout.
Then, $\structout \prec_\Gamma \structout'$ iff there exists some symbol $S \in \vocout$ and domain element tuple $\bar{d}$ such that $\bar{d} \not \in S^{\structout}$, $\bar{d} \in S^{\structout'}$, and for all $\bar{d}' \prec_\domain \bar{d}$ it holds that $\bar{d}' \in S^{\structout} \Leftrightarrow \bar{d}' \in S^{\structout'}$, and for all $S' \prec_{\vocout} S$, it holds that $S'^{\structout}=S'^{\structout'}$.
%
For the remainder of this section, we leave the order over $\vocout$ implicit, but explicitly state the order $\preceq_\domain$ over \domain, as this turns out to be important.
Given a model expansion problem with symmetry $\sigma$ and a lexicographical order over $\Gamma_\domain^\vocout$ with $\preceq_\domain$ as \domain-order, we use $\lex{\sigma}$ to refer to the logical formula encoding the lex-leader constraint for $\sigma$. Efficient encodings of the lex-leader constraint into formulas are well-known~\cite{Sak09HBSAT}.

\begin{example}[Example~\ref{ex:graph_coloring_start} continued]
Let $t \prec_\domain u \prec_\domain v \prec_\domain w \prec_\domain r \prec_\domain g \prec_\domain b$ and $A=\{\argpos{C}{1},\argpos{Color}{0}\}$.
For $MX(\theorygc,\structgcin)$, 
the local domain symmetry $\locdomsym{(r~g)}{A}$  is broken by the lex-leader constraint $\lex{\locdomsym{(r~g)}{A}}$, which informally implies that for each vertex $v$, if all vertices $v' \prec_{\domain} v$ are not colored by $r$ or by $g$, then $v$ cannot be colored with $r$. 
Amongst others, this constraint cuts away $\vocout$-structures 
that color $t$ with $r$.
\demo\end{example}

Note that lex-leader constraints are constructed for individual
symmetries. In general, to obtain a complete symmetry breaking
constraint for a symmetry group $\mathbb{G}$, one needs to post
$\lex{\sigma}$ for each $\sigma \in \mathbb{G}$. As symmetry groups can contain a factorial amount of symmetries this is infeasible, e.g., in the case of subdomain interchangeability.
Instead, the standard approach is \emph{partial symmetry breaking}, where
$\lex{\sigma}$ is posted for a minimal set of generators $\sigma$ of
$\mathbb{G}$~\cite{Shatter}. 
Partial symmetry breaking is feasible, but does not guarantee that $\mathbb{G}$ is broken completely, leaving symmetrical parts of the search space open
to a search engine.

For instance, for a subdomain interchangeability group $\locdomintch{\delta}{A}$, a minimal set of generator symmetries is $\{\locdomsym{(d~s(d))}{A} \mid d, s(d) \in \delta\}$, where $s(d)$ is the successor of $d$ in $\delta$ according to $\preceq_\domain$.
Other minimal generator sets exist as well, e.g., $\{\locdomsym{(d_0~d)}{A} \mid d \in \delta, d\neq d_0\}$ for a fixed $d_0 \in \delta$.
However, the choice of the generator set influences the power of the symmetry breaking formula.
For subdomain interchangeability groups $\mathbb{G}$, choosing the right generator set can guarantee that the lex-leader constraints used in partial symmetry breaking are actually complete for $\mathbb{G}$:

\begin{theorem}
\label{thm:complete_breaking}
Let $MX(\theory,\structin)$ be a model expansion problem, $\delta$
an $A$-interchangeable subdomain,
$\preceq_\domain$ a total order on domain \domain and $s(d)$ the successor of $d$ in $\delta$ according to $\preceq_\domain$.
If $A$ contains at most one argument position $\argpos{S}{i}$ for each symbol $S \in \vocout$, then the conjunction of lex-leader constraints
\[ \{ \lex{\locdomsym{(d~s(d))}{A}} \mid d, s(d) \in \delta \} \]
is a complete symmetry breaking constraint for the subdomain
interchangeability group $\locdomintch{\delta}{A}$.
\end{theorem}

A strongly related result is that when constructing a relation $R \subseteq \domain_1 \times \ldots \times \domain_n$ for which exactly one dimension $D_i$ contains interchangeable values, an efficient lex-leader constraint exists that completely breaks the resulting symmetry~\cite{dam/shlyakter2007}.
Theorem~\ref{thm:complete_breaking} can be seen as a conversion of this result to a model expansion context with local domain interchangeability.

Intuitively, Theorem~\ref{thm:complete_breaking} states that local
domain interchangeability is completely broken by a linear number of lex-leader
constraints if the set of argument positions contains at most one
argument position for each output symbol. These lex-leader constraints
$\lex{\locdomsym{(d~s(d))}{A}}$ are based on swaps $(d~s(d))$
of two consecutive domain elements over the chosen domain
ordering. Note that lex-leader constraints based on swaps of
non-consecutive domain elements, e.g., $\{\locdomsym{(d_0~d)}{A} \mid d \in \delta, d\neq d_0\}$ for a fixed $d_0$, do not have this property~\cite{cspsat/DevriendtBB14}.

\begin{example}[Example~\ref{ex:graph_coloring_start} continued]
Given the graph coloring problem $MX(\theorygc,\structgcin)$, let $A=\{\argpos{C}{1},\argpos{Color}{0}\}$ and $r \prec_\domain g \prec_\domain b$. $\locdomintch{\{r,g,b\}}{A}$ is a subdomain interchangeability group of $MX(\theorygc,\structgcin)$. 
Since $A$ contains only one argument position for the symbol $Color$, it is completely broken by 
\begin{flalign*}
 \qquad\qquad\qquad\qquad\qquad\qquad\quad&\lex{\locdomsym{(r~g)}{A}} \land \lex{\locdomsym{(g~b)}{A}}&\demo 
 \end{flalign*}
\end{example}

The finite model generation system \SEM{} also breaks this type of symmetry completely, by way of \emph{dynamically} avoiding symmetrical decisions during search.~\cite{Zhang95sem}
The more recent model generator \kodkod{}~\cite{tacas/TorlakJ07} breaks symmetry statically by posting lex-leader constraints from~\citeN{Shatter} for global domain symmetry.
Although~\citeN{tacas/TorlakJ07} do not mention any completeness result, experiments with a pigeonhole encoding in \kodkod{} indicate that it uses the right set of generator symmetries to completely break all pigeon and hole interchangeability symmetry.

\section{Symmetry detection}
\label{sec:detection}
In this section, we give two local domain symmetry detection algorithms for model expansion problems.
The first detects generators of a local domain symmetry group, the second derives interchangeable subdomains.
Both approaches work on a first-order level, avoiding the need to \emph{ground} the model expansion problem to a propositional counterpart.
Both algorithms are based on Theorem~\ref{thm:locdomsymconditionmx}, which conditions argument position set $A$ to be connectively closed under decomposition theory $\structinstar$.
To find such $A$, one simply constructs a partition of $\theorystar$'s argument positions.
Using a disjoint-set data structure\footnote{\label{note:disjointset}\url{en.wikipedia.org/wiki/Disjoint-set_data_structure}},
the computational cost to find $A$ is linear in the size of $\theory$.
In the following subsections, we assume a set of argument positions $A$ satisfying the connectedness condition is available, leaving only the concern of finding an appropriate domain permutation $\pi$ (Section \ref{sec:locdomsymdetect}) or interchangeable subdomain $\delta$ (Section \ref{sec:subd_intch_detection}).

\subsection{Local domain symmetry detection}
\label{sec:locdomsymdetect}
%
Our approach follows other symmetry detection techniques~\cite{2002Aloul,drtiwa11a} by converting the symmetry detection problem to a \emph{graph automorphism} detection problem.
An \emph{automorphism} of a graph is a permutation $\tau$ of its vertices such that each vertex pair $(v,u)$ forms an edge iff $(\tau(v),\tau(u))$ forms an edge. If the graph is colored, then each vertex $v$ must have the same color as $\tau(v)$.

This existing work encodes a propositional theory into a graph, which we call the \emph{detection graph}.
If the detection graph is well-constructed, its automorphism group corresponds to a symmetry group of the propositional theory.
Tools such as \saucy{}~\cite{saucy} then are employed to derive generators for the detection graph's automorphism group, which in turn are converted to symmetry generators for the propositional theory.

Our approach differs by not encoding a propositional theory into the detection graph, but an input structure and a set of argument positions, as these are all we need to detect local domain symmetry.
Formally, given a structure $\struct$ and an argument position set $A$, we construct an undirected colored graph whose automorphisms correspond to domain permutations $\pi$ such that $\locdomsym{\pi}{A}(\struct)=\struct$ -- satisfying the second condition of Theorem~\ref{thm:locdomsymconditionmx}.

\begin{definition}[Domain permutation graph]
Let $\struct$ be a $\voc$-structure with domain $\domain$ and $A$ a set of argument positions.
The \emph{domain permutation graph} $\dpg{\struct}{A}$ for $\struct$ and $A$ is an undirected colored graph with labeled vertices $V$, edges $E$ and color function $C$ that satisfies the following requirements:

$V$ is partitioned into three subsets: $DE$ (domain element vertices), $AP$ (argument position vertices) and $IT$ (interpretation tuple vertices).
$DE$ contains a vertex labeled $d$ for each $d \in \domain$.
$AP$ contains $k+1$ vertices labeled $\{\argnode{d}{i} \mid i \in [0..k]\}$ for each $d \in \domain$, with $k$ the maximum arity of symbols in $\voc$.
$IT$ contains a vertex labeled $S(\bar{d})$ for each tuple $\bar{d} \in S^{\struct}$ with $S^\struct \in \struct$.

$E$ consists only of edges between $DE$ and $AP$, and between $AP$ and $IT$.
An $AP$ vertex labeled $\argnode{d}{i}$ is connected to a $DE$ vertex $e$ iff $d=e$.
An $IT$ vertex labeled $S(\ldots,d_i,\ldots)$ is connected to an $AP$ vertex $\argnode{e}{j}$ iff $d=e$, $i=j$ and $\argpos{S}{i} \in A$.

Vertices from different partitions have different colors.
All $DE$ vertices have the same color.
Two $AP$ vertices labeled $\argnode{d}{i}$ and $\argnode{e}{j}$ have the same color iff $i=j$.
Two $IT$ vertices labeled $S(d_1,\ldots,d_n)$ and $R(e_1,\ldots,e_n)$ have the same color iff $S=R$ and $d_i=e_i$ for all $i$ such that $\argpos{S}{i} \not \in A$.
\end{definition}



The intuition behind the domain permutation graph $\dpg{\struct}{A}$ is that a permutation of its $DE$ vertices corresponds to a domain permutation $\pi$, a permutation of its $IT$ vertices corresponds to a permutation of domain element tuples in interpretations in $\struct$, and the $AP$ vertices and vertex coloring serve to link $DE$ and $IT$ in such a way that Definition~\ref{def:inducedstructuretransformation} is preserved for automorphisms.

\begin{theorem}
\label{thm:domainpermutationgraph}
Let $\struct$ be a $\voc$-structure with domain $\domain$ and $A$ a set of argument positions.
There exists a bijection between the automorphism group of the domain permutation graph \dpg{\struct}{A} and the group of domain permutations $\pi$ such that $\locdomsym{\pi}{A}(\struct)=\struct$. This bijection maps an automorphism $\tau$ to domain permutation $\pi$ iff $\tau(d)=\pi(d)$ for all $DE$ vertices (equated with domain elements) $d$.
\end{theorem}

\begin{figure}
\label{fig:gcdpg}
\begin{tikzpicture}[-,scale=0.9]
  \tikzstyle{denode}=[regular polygon,regular polygon sides=3,minimum size=22pt,inner sep=0pt,draw]
  \tikzstyle{argnode0}=[diamond,minimum size=22pt,inner sep=0pt,draw]
  \tikzstyle{argnode1}=[circle,minimum size=22pt,inner sep=0pt,draw]
  \tikzstyle{argnode2}=[regular polygon,minimum size=22pt,inner sep=0pt,draw]
  \tikzstyle{tuplenode1}=[rectangle,minimum width=55pt, minimum height=20pt,inner sep=0pt,draw]
  \tikzstyle{tuplenode2}=[diamond,minimum size=22pt,inner sep=0pt,draw]

  \foreach \name/\x in {t/0, u/2.5, v/5, w/7.5}
    \node[denode] (\name) at (\x-1,-1) {$\name$};
  \foreach \name/\x in {t/0, u/2.5, v/5, w/7.5}
    \node[argnode2] (\name2) at (\x-1,-2) {$\argnode{\name}{2}$};
  \foreach \name/\x in {t/1, u/3.5, v/6, w/8.5}
    \node[argnode1] (\name1) at (\x-1,-2) {$\argnode{\name}{1}$};
  \foreach \name/\x in {t/1, u/3.5, v/6, w/8.5}
    \node[argnode0] (\name0) at (\x-1,-1) {$\argnode{\name}{0}$};

  \foreach \name/\x/\y in {r/10/-1, g/12/-1, b/11/-3}
    \node[denode] (\name) at (\x-1.3,\y) {$\name$};
  \foreach \name/\x/\y in {r/9/-2, g/11/-2, b/10/-4}
    \node[argnode2] (\name2) at (\x-0.3,\y) {$\argnode{\name}{2}$};
  \foreach \name/\x/\y in {r/10/-2, g/12/-2, b/11/-4}
    \node[argnode1] (\name1) at (\x-0.3,\y) {$\argnode{\name}{1}$};
  \foreach \name/\x/\y in {r/10/-1, g/12/-1, b/11/-3}
    \node[argnode0] (\name0) at (\x-0.3,\y) {$\argnode{\name}{0}$};

  \foreach \pred/\arg/\argg/\x in {Edge/t/u/0, Edge/u/v/2.5, Edge/v/w/5, Edge/w/t/7.5}
    \node[tuplenode1] (\pred\arg\argg) at (\x-0.5,-4) {$\pred_1(\arg,\argg)$};

  \foreach \from/\to in {t/0,t/1,t/2,u/0,u/1,u/2,v/0,v/1,v/2,w/0,w/1,w/2,r/0,r/1,r/2,g/0,g/1,g/2,b/0,b/1,b/2}
    \draw (\from) -- (\from\to);

  \foreach \from/\to in {t1/tu,t2/wt,u1/uv,u2/tu,v1/vw,v2/uv,w1/wt,w2/vw}
    { \draw (\from) -- (Edge\to);}
\end{tikzpicture}
\caption{Domain permutation graph \dpg{\structgcinstar}{A} with $A=\{\argpos{Edge_1}{1},\argpos{Edge_1}{2},\argpos{Color}{1},\argpos{x_1}{0},\argpos{y_1}{0},\argpos{x_3}{0}\}$. Each shape denotes a unique color, so vertices with the same shape have the same color.}
\end{figure}
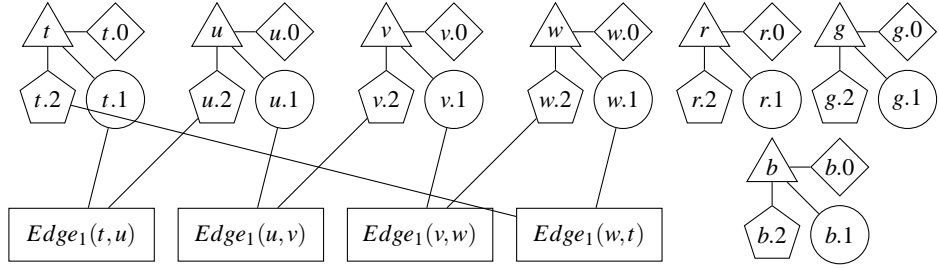

\begin{example}[Example~\ref{ex:graph_coloring_7} continued]
Using argument position set $A=\{\argpos{Edge_1}{1},\argpos{Edge_1}{2},\argpos{Color}{1},\argpos{x_1}{0},\argpos{y_1}{0},\argpos{x_3}{0}\}$ (which is connectively closed under $\theorygcstar$) and input structure $\structgcinstar$, the domain permutation graph \dpg{\structgcinstar}{A} is illustrated in Figure~\ref{fig:gcdpg}.
The automorphism group of \dpg{\structgcinstar}{A} corresponds to the group of induced structure transformations $\locdomsym{\pi}{A}$ such that $\locdomsym{\pi}{A}(\structgcinstar)=\structgcinstar$.
As a result, its automorphism group corresponds to a local domain symmetry group of $MX(\theorygc,\structgcin)$.
E.g., $\locdomsym{(t~u~v~w)}{A}$ corresponds to an automorphism that permutes the four left-most groups of five vertices, and $\locdomsym{(b~g)}{A}$ to an automorphism that swaps the two right-most groups of four vertices.
\demo\end{example}

Let $k$ be the largest arity of a symbol in $\struct$ for a domain permutation graph $\dpg{\struct}{A}$.
The size of $DE$ is $|\domain|$, the size of $AP$ is $(k+1)|\domain|$, and the size of $IT$ is $|\struct|$, which is $O(|\domain|^k)$.
Thus, the total number of nodes is $O(k|\domain|+|\domain|^k)$.
There are $(k+1)|\domain|$ edges between $DE$ and $AP$, and, if all argument positions over some symbol $S/k$ occur in $A$, then there are $O(k|\struct|)=O(k|\domain|^k)$ edges between $AP$ and $IT$.
Thus, the total number of edges is $O(k|\domain|^k)$.

Note that the size of \dpg{\struct}{A} does not depend on the size of the theory of the model expansion problem.
This is a major advantage compared to automorphism-based symmetry detection on ground theories, as the detection graph grows linearly with the ground theory~\cite{drtiwa11a}, which is typically much larger than the input structure.

\subsection{Subdomain interchangeability detection}
\label{sec:subd_intch_detection}
While the previous subsection detects local domain symmetry generators individually, it is not clear what type of symmetry group they form.
To optimally construct symmetry breaking constraints for symmetry groups, we need to detect subdomain interchangeability as well.
Then, by Theorem~\ref{thm:complete_breaking}, we will often be able to break subdomain interchangeability groups completely with a set of lex-leader constraints linear in $|\domain|$.

Given a model expansion problem $MX(\theory,\structin)$ with decomposition $MX(\theorystar,\structinstar)$ and a set of argument positions $A$ connectively closed under $\theorystar$, 
the task at hand is to find a subdomain $\delta \subseteq \domain$ such that for each permutation $\pi$ over $\delta$, $\locdomsym{\pi}{A}(\structinstar)=\structinstar$.
If so, Theorem~\ref{thm:locdomsymconditionmx} guarantees $\locdomsym{\pi}{A}$ to be a symmetry of $MX(\theory,\structin)$, which makes $\delta$ an $A$-interchangeable subdomain.

The actual algorithm finds a partition $\Delta$ of \domain, such that each $\delta \in \Delta$ is $A$-interchangeable.
The idea is based on the fact that the permutation group of a set is generated by swaps of two elements of the set.
As such, if we know for each pair $d_1,d_2 \in \domain$ whether $\locdomsym{(d_1~d_2)}{A}(\structinstar)=\structinstar$, it is straightforward to construct the partition $\Delta$.
The resulting symmetry detection algorithm is simple: for each pair of domain elements $d_1,d_2 \in \domain$, check whether $\locdomsym{(d_1~d_2)}{A}(\structinstar)=\structinstar$.
When using a disjoint-set data structure\footref{note:disjointset} to keep track of the partition $\Delta$, the complexity of this algorithm is $O(|\domain|^2|\structinstar|)$.
The algorithm can be optimized by exploiting transitivity, domain element occurrence counting or unary symbols partitioning the domain, but this does not improve the worst-case complexity.

\begin{example}[Example~\ref{ex:graph_coloring_7} continued]
Given argument position set $A=\{\argpos{C_1}{1},\argpos{Color}{0}\}$ (which is connectively closed under $\theorygcstar$), we detect $A$-interchangeable domains by checking whether the (only) input symbol $C_1$ has the same interpretation in $\locdomsym{(d_1~d_2)}{A}(\structinstar)$ as in $\structinstar$ for combinations of $d_1,d_2 \in \{t,u,v,w,r,g,b\}$.
For $(d_1,d_2) \in \{(t,u),(u,v),(v,w),(r,g),(g,b)\}$ this is indeed the case. For $(d_1,d_2)=(w,r)$ this is not the case, so the $A$-interchangeable sets are $\{t,u,v,w\}$ and $\{r,g,b\}$.
\demo\end{example}


\section{Experiments}
\label{sec:experiments}
Based on the theory and algorithms presented in this paper, we implemented symmetry exploitation in the model expansion inference of the \idp system~\mycite{IDP}. 
\idp is a \emph{knowledge base system} where knowledge about a problem can be modelled in \fodot, a rich extension of first-order logic~\mycite{fodot}.
Our implementation (including source code) is available online\footnote{\label{note:bbexperiments}\url{bitbucket.org/krr/fo-sym-experiments}}, and is incorporated in \idp version 3.6.0 and up.
Our implementation makes use of \saucy{} version 3 to solve the graph automorphism component of symmetry detection, and constructs symmetry breaking formulas based on the lex-leader encoding of \citeN{Shatter}.

We compare this implementation with the ASP system \clasp{}~\cite{gekakasc14b} version 3.1.4, using version 4.5.4 of the ASP grounder \gringo{} to generate ground answer set programs.
For \clasp{}, the symmetry breaking preprocessor \sbass{} has been developed~\cite{drtiwa11a}.
\sbass{} takes a ground answer set program, encodes it to a detection graph, uses \saucy{} to solve the automorphism detection problem, converts \saucy{}'s output to permutations of propositional atoms that induce symmetries, and constructs symmetry breaking constraints following \citeN{Shatter}.

Our experiment uses four different system configurations: \idp{} and \clasp{} refer to both systems without symmetry breaking, \idpsym{} refers to \idp{} extended with the techniques described in this paper, and \sbass{} refers to \clasp{} coupled with the symmetry breaking preprocessor.

This experiment can only broadly compare the \idp{} and \clasp{} configurations, as both systems use similar but ultimately different techniques to solve the model expansion problem.
Our main interest is to investigate the types of symmetry detected, the overhead needed to detect those, and the relative speedup gained when activating symmetry algorithms for both systems.
We expect that \idpsym{}, compared to \sbass{}, has less symmetry detection overhead, as \idpsym{} detects symmetry on the first-order level instead of on the ground level.
E.g., the structure information present in a set of connectively closed argument positions can be derived with a syntactical check on the first-order theory, but this information is lost after grounding.
As a result, we expect \idpsym{}'s detection graph to be smaller, or even non-existent.\footnote{\label{note:idpsymopt}\idpsym{} does not construct the detection graph if the only generators it will detect are due to subdomain interchangeability. 
Given an argument position set $A$, this is the case if for each symbol $S$, $A$ contains at most one argument position over $S$.}
Also, we expect a larger relative speedup for \idpsym{} than for \sbass{} on problems with a lot of subdomain interchangeability, as only \idpsym{} detects and completely breaks this type of symmetry.
However, as mentioned in Section \ref{sec:more_symmetry}, \idpsym{}'s detected symmetry group might be smaller than \sbass{}'s, as not all symmetry properties of a problem can be captured by our notion of local domain symmetry.

Our benchmark set consists of four problem families: \holes{}, \crew{}, \graceful{} and \nqueens{}. 
\holes{} is a set of 16 unsatisfiable pigeonhole instances where $n$ pigeons must be placed in $n-1$ different holes. 
The pigeons and holes are indistinguishable, leading to subdomain interchangeability symmetry groups.
\crew{} is a set of 42 unsatisfiable airline crew scheduling instances, where optimality has to be proven for a minimal crew assignment given a moderately complex flight plan.
The instances are generated by hand, with the number of crew members ranging from $5$ to $25$.
Crew members have different attributes, but depending in the instance, multiple crew members exist with exactly the same attribute set, leading to subdomain interchangeability symmetry.
\graceful{} consists of 60 graceful graph instances, taken from 2013's ASP competition. 
These instances require to label a graph's vertices and edges such that all vertices have a different label, all edges have a different label, and each edge's label is the difference of the labels of the vertices it connects.
The labels used are $\{0,1,\dots,n\}$, with $n$ the number of edges.
Any symmetry exhibited by the input graph is present, as well as a symmetry mapping each vertex' label $l$ to $n-l$.
\nqueens{} is one N-Queens instance trying to fit 200 queens on a $200$ by $200$ chessboard so that no queen threatens another.
The symmetry present in \nqueens{} is due to the rotational and reflective symmetries of the chessboard.

The available resources were 6GB RAM and 1000s timeout on an Intel\textsuperscript{\textregistered} Xeon\textsuperscript{\textregistered} E3-1225 CPU with Ubuntu 14.04 Linux kernel 3.13 as operating system.
\fodot{} and ASP specifications, instances and detailed experimental results are available online.\footref{note:bbexperiments}
Table~\ref{tbl:results} summarizes the results.

\begin{table}[htb]
\centering
\begin{tabular}{ l | c | cccc | c | ccccc | }			
  & \multicolumn{1}{l|}{\clasp{}} & \multicolumn{4}{c|}{\sbass{}} &  \multicolumn{1}{c|}{\idp}  & \multicolumn{5}{c|}{\idpsym} \\
  & $\#$ & $\#$ & $t$ & $V$ & $\pi$ & $\#$ & $\#$ & $t$ & $V$ & $\pi$ & $\delta$ \\
\hline
\holes{} (16)    & 8  & 11 & 50.9 & 48814   & 43.5 & 8  & 16 & 0.0  & 0~\footref{note:idpsymopt} & 0   & 2   \\
\crew{} (42)    & 32  & 36 & 0.0 & 1722   & 7.8 & 28  & 39 & 0.0  & 0~\footref{note:idpsymopt} & 0   & 4.1   \\
\graceful{} (60) & 33 & 20 & 0.7  & 127860   & 5.5  & 26 & 13 & 0.4  & 15201 & 5.4 & 0.6 \\
\nqueens{} (1)   & 1  & 1  & 76.2 & 9357802 & 2    & 1  & 1  & 6.8 & 0~\footref{note:idpsymopt} & 0   & 0   
\end{tabular}
\caption{Experimental results of \clasp{}- and \idp{}-based solvers with and without symmetry breaking. $\#$ represents the number of solved instances, $t$ the average symmetry detection time in seconds, $V$ the average number of vertices in the detection graph, $\pi$ the average number of symmetry generators detected by \saucy{}, and $\delta$ the average number of interchangeable subdomains detected.}
\label{tbl:results}
\end{table}

When analyzing the results on \holes{}, it is clear that plain \clasp{} and \idp{} get lost in symmetric parts of the search tree, solving only $8$ instances (up to $12$ pigeons).
\sbass{} can only solve three more instances (up to $15$ pigeons), as the derived symmetry generators do not suffice to construct strong symmetry breaking constraints.
These results are consistent with~\citeN{drtiwa11a}.
\idpsym{} detects the pigeon and hole interchangeable subdomains, and its complete symmetry breaking constraints allow all $16$ instances to be solved (up to $100$ pigeons).
As far as symmetry detection time goes, unlike \sbass{}, \idpsym{} has negligible detection overhead.

The results on \crew{} are similar to \holes{} but less outspoken.
The reason is that even though there are more subdomain interchangeability groups, the subdomains are a lot smaller, incurring less symmetry overhead.
As a result, \idpsym{} only enjoys a small advantage over \sbass{}, but does reverse the situation where pure \clasp{} outperforms pure \idp{}.
Concerning symmetry detection, \idpsym{} has to  analyze the input structure before deriving any subdomain interchangeability groups, which contrasts with the trivially interchangeable pigeons and holes in \holes{}.
Nonetheless, \idpsym{} solves this task in the blink of an eye, as does \sbass{}.

Continuing with \graceful{}, it is striking that the number of solved instances is \emph{reduced} by symmetry breaking.
Upon closer inspection, this is only the case for satisfiable instances.
For unsatisfiable \graceful{} instances, \sbass{} solves two more than \clasp{}, and \idpsym{} solves three more than \idp{}.
This discrepancy is not uncommon, as static symmetry breaking reduces the search space by removing possibly easy-to-find solutions.
These results are also consistent with those reported by~\citeN{drtiwa11a}.
Looking at the number of symmetry generators detected, both \sbass{} and \idpsym{} detect about the same number of symmetry generators, indicating that they detect the same symmetry group.
Note that \idpsym{} also detects a few subdomain interchangeability groups, apparently present in the input graph.
As far as symmetry detection overhead goes, \sbass{} is slower than \idpsym{}.
This is not surprising, as \sbass{}' detection graph is almost an order of magnitude larger than \idpsym{}'s.
We conclude that for \graceful{}, \idpsym{} detects the same symmetry group as \sbass{}, though with less overhead.

Lastly, for \nqueens{}, \idpsym{} cannot detect the geometric symmetries of the chessboard, as this type of symmetry does not fit the definition of local domain symmetry. 
For instance, a square $(i,j)$ on the chess board is diagonally reflected to square $(j,i)$, while square $(i,k)$ is reflected to $(k,i)$. Domain element $i$ is mapped to both $j$ and $k$ at position $0$, violating the local domain symmetry requirement that it stems from a domain \emph{permutation}.
\sbass{} can detect this type of symmetry, as it detects permutations of ground atoms instead of domain elements. However, note the significant overhead incurred, as the detection graph is huge.

We conclude that our approach has very low symmetry detection overhead, due to a smaller or non-existent detection graph. Moreover, by completely breaking subdomain interchangeability, we significantly increase the number of solved instances.
However, not all symmetry present in the problem set is detected by our approach.

\section{Conclusion}
\label{sec:conclusion}
We presented the notion of local domain symmetry for model expansion problems, which manifests itself on the first-order level.
We gave a completeness result on the strength of symmetry breaking constraints for a special case of local domain symmetry, and we posted syntactical conditions to efficiently detect symmetry from a model expansion specification.
Our experiment highlights the strengths and weaknesses of our approach. We have a very low symmetry detection overhead and we give symmetry breaking completeness guarantees for local domain interchangeability that are effective in practice. However, we cannot detect some forms of symmetry.

It is worth mentioning that local domain symmetry is not limited to pure classical logic; it is straightforward to extend our work to cardinalities, types or arithmetic.
Similarly, logic programs under stable or well-founded semantics have symmetry properties induced by permutations of domain elements (or Herbrand constants) and sets of argument positions.
Our work easily transfers to these domains. In fact, our implementation in \idp{} already supports such extensions.

Investigating which types of symmetry fall outside our formalism, and inventing ways to detect and exploit these types of symmetry is interesting future work.
One idea is that not only permutations of the domain lead to symmetry, but permutations on (argument positions of) symbols in the vocabulary do as well.
We suspect there is also room for improvement on the symmetry breaking front. Limiting the size of individual symmetry breaking constraints is known to improve their performance.
Also, to avoid deteriorating performance on satisfiable instances, one should look into dynamic symmetry breaking approaches, as these typically do not cut away solutions a priori.

\section{Acknowledgements}
\label{sec:acknowledgements}
This research was supported by the project GOA 13/010 Research Fund
KU Leuven and projects G.0489.10, G.0357.12 and G.0922.13 of FWO
(Research Foundation - Flanders). 
Bart Bogaerts is supported by the Finnish Center of Excellence in Computational
Inference Research (COIN) funded by the Academy of Finland (grant \#251170).


\bart{Als je de bibligrafie korter maakt (bijvorobeeld branch BADREFS\_REBASE gebruiken van idp-latex, dan zit je binnen de paginalimiet}
\bibliographystyle{acmtrans}
\bibliography{idp-latex/krrlib,custom}

\begin{thebibliography}{}

\bibitem[\protect\citeauthoryear{Aloul, Ramani, Markov, and Sakallah}{Aloul
  et~al\mbox{.}}{2002}]{2002Aloul}
{\sc Aloul, F.}, {\sc Ramani, A.}, {\sc Markov, I.}, {\sc and} {\sc Sakallah,
  K.} 2002.
\newblock Solving difficult {SAT} instances in the presence of symmetry.
\newblock In {\em Design Automation Conference, 2002. Proceedings. 39th}.
  731--736.

\bibitem[\protect\citeauthoryear{Aloul, Sakallah, and Markov}{Aloul
  et~al\mbox{.}}{2006}]{Shatter}
{\sc Aloul, F.~A.}, {\sc Sakallah, K.~A.}, {\sc and} {\sc Markov, I.~L.} 2006.
\newblock Efficient symmetry breaking for {B}oolean satisfiability.
\newblock {\em IEEE Transactions on Computers\/}~{\em 55,\/}~5, 549--558.

\bibitem[\protect\citeauthoryear{Audemard and Benhamou}{Audemard and
  Benhamou}{2002}]{Audemard02reasoningby}
{\sc Audemard, G.} {\sc and} {\sc Benhamou, B.} 2002.
\newblock Reasoning by symmetry and function ordering in finite model
  generation.
\newblock In {\em Automated Deduction - CADE-18, 18th International Conference
  on Automated Deduction, Copenhagen, Denmark, July 27-30, 2002, Proceedings},
  {A.~Voronkov}, Ed. Lecture Notes in Computer Science, vol. 2392. Springer,
  226--240.

\bibitem[\protect\citeauthoryear{Claessen and S\"{o}rensson}{Claessen and
  S\"{o}rensson}{2003}]{Claessen03newtechniques}
{\sc Claessen, K.} {\sc and} {\sc S\"{o}rensson, N.} 2003.
\newblock {New Techniques that Improve MACE-style Model Finding}.
\newblock In {\em Workshop on Model Computation (MODEL)}.

\bibitem[\protect\citeauthoryear{{De Cat}, Bogaerts, Bruynooghe, Janssens, and
  Denecker}{{De Cat} et~al\mbox{.}}{2016}]{WarrenBook/DeCatBBD14}
{\sc {De Cat}, B.}, {\sc Bogaerts, B.}, {\sc Bruynooghe, M.}, {\sc Janssens,
  G.}, {\sc and} {\sc Denecker, M.} 2016.
\newblock Predicate logic as a modelling language: The {IDP} system.
\newblock {\em CoRR\/}~{\em abs/1401.6312v2}.

\bibitem[\protect\citeauthoryear{Denecker and Ternovska}{Denecker and
  Ternovska}{2008}]{tocl/DeneckerT08}
{\sc Denecker, M.} {\sc and} {\sc Ternovska, E.} 2008.
\newblock A logic of nonmonotone inductive definitions.
\newblock {\em ACM Trans. Comput. Log.\/}~{\em 9,\/}~2 (Apr.), 14:1--14:52.

\bibitem[\protect\citeauthoryear{Devriendt, Bogaerts, and Bruynooghe}{Devriendt
  et~al\mbox{.}}{2014}]{cspsat/DevriendtBB14}
{\sc Devriendt, J.}, {\sc Bogaerts, B.}, {\sc and} {\sc Bruynooghe, M.} 2014.
\newblock {BreakIDGlucose}: On the importance of row symmetry in {SAT}.
\newblock In {\em Proceedings of the Fourth International Workshop on the
  Cross-Fertilization Between {CSP} and {SAT} ({CSPSAT})}.

\bibitem[\protect\citeauthoryear{Drescher, Tifrea, and Walsh}{Drescher
  et~al\mbox{.}}{2011}]{drtiwa11a}
{\sc Drescher, C.}, {\sc Tifrea, O.}, {\sc and} {\sc Walsh, T.} 2011.
\newblock Symmetry-breaking answer set solving.
\newblock {\em IA Communications\/}~{\em 24,\/}~2, 177--194.

\bibitem[\protect\citeauthoryear{Enderton}{Enderton}{2001}]{Enderton01}
{\sc Enderton, H.~B.} 2001.
\newblock {\em A Mathematical Introduction To Logic\/}, Second ed.
\newblock Academic Press.

\bibitem[\protect\citeauthoryear{Flener, Frisch, Hnich, Kiziltan, Miguel,
  Pearson, and Walsh}{Flener et~al\mbox{.}}{2002}]{row_column_sym_csp}
{\sc Flener, P.}, {\sc Frisch, A.~M.}, {\sc Hnich, B.}, {\sc Kiziltan, Z.},
  {\sc Miguel, I.}, {\sc Pearson, J.}, {\sc and} {\sc Walsh, T.} 2002.
\newblock Breaking row and column symmetries in matrix models.
\newblock In {\em Principles and Practice of Constraint Programming - CP 2002},
  {P.~Hentenryck}, Ed. LNCS, vol. 2470. Springer Berlin Heidelberg, 462--477.

\bibitem[\protect\citeauthoryear{Gebser, Kaminski, Kaufmann, and Schaub}{Gebser
  et~al\mbox{.}}{2014}]{gekakasc14b}
{\sc Gebser, M.}, {\sc Kaminski, R.}, {\sc Kaufmann, B.}, {\sc and} {\sc
  Schaub, T.} 2014.
\newblock \textit{Clingo} = {ASP} + control: Preliminary report.
\newblock In {\em Technical Communications of the Thirtieth International
  Conference on Logic Programming (ICLP'14)}, {M.~Leuschel} {and}
  {T.~Schrijvers}, Eds. Vol. 14(4-5).
\newblock Online Supplement.

\bibitem[\protect\citeauthoryear{Gent, Petrie, and Puget}{Gent
  et~al\mbox{.}}{2006}]{gent2006symmetry}
{\sc Gent, I.~P.}, {\sc Petrie, K.~E.}, {\sc and} {\sc Puget, J.-F.} 2006.
\newblock Symmetry in constraint programming.
\newblock {\em Handbook of Constraint Programming\/}~{\em 10}, 329--376.

\bibitem[\protect\citeauthoryear{Katebi, Sakallah, and Markov}{Katebi
  et~al\mbox{.}}{2010}]{saucy}
{\sc Katebi, H.}, {\sc Sakallah, K.~A.}, {\sc and} {\sc Markov, I.~L.} 2010.
\newblock Symmetry and satisfiability: An update.
\newblock In {\em SAT}, {O.~Strichman} {and} {S.~Szeider}, Eds. LNCS, vol.
  6175. Springer, 113--127.

\bibitem[\protect\citeauthoryear{Sakallah}{Sakallah}{2009}]{Sak09HBSAT}
{\sc Sakallah, K.~A.} 2009.
\newblock {\em {Symmetry and Satisfiability}}. Frontiers in Artificial
  Intelligence and Applications, vol. 185.
\newblock IOS Press, Chapter~10, 289--338.

\bibitem[\protect\citeauthoryear{Shlyakhter}{Shlyakhter}{2007}]{dam/shlyakter2007}
{\sc Shlyakhter, I.} 2007.
\newblock Generating effective symmetry-breaking predicates for search
  problems.
\newblock {\em Discrete Appl. Math.\/}~{\em 155,\/}~12 (June), 1539--1548.

\bibitem[\protect\citeauthoryear{Torlak and Jackson}{Torlak and
  Jackson}{2007}]{tacas/TorlakJ07}
{\sc Torlak, E.} {\sc and} {\sc Jackson, D.} 2007.
\newblock Kodkod: A relational model finder.
\newblock In {\em TACAS}, {O.~Grumberg} {and} {M.~Huth}, Eds. LNCS, vol. 4424.
  Springer, 632--647.

\bibitem[\protect\citeauthoryear{Walsh}{Walsh}{2012}]{walsh_recent_results_2012}
{\sc Walsh, T.} 2012.
\newblock Symmetry breaking constraints: Recent results.
\newblock {\em CoRR\/}~{\em abs/1204.3348}.

\bibitem[\protect\citeauthoryear{Zhang and Zhang}{Zhang and
  Zhang}{1995}]{Zhang95sem}
{\sc Zhang, J.} {\sc and} {\sc Zhang, H.} 1995.
\newblock Sem: A system for enumerating models.
\newblock In {\em Department of Philosophy University of Wisconsin-Madison
  Mathematics and Computer Science}. 298--303.

\end{thebibliography}

\newpage
\appendix
\section{Proofs}\label{app:proofs}

\noindent
\textit{Theorem~\ref{thm:locdomsymcondition}}\\
\noindent
Let \voc be a vocabulary, \theory a theory over $\voc$, $\pi$ a domain permutation and $A$ a set of argument positions. If $A$ is connectively closed under \theory, then $\locdomsym{\pi}{A}$ is a local
domain symmetry of \theory.
\begin{proof}
To prove this theorem, we prove the following consecutive claims for each \voc-structure $\struct$. 
Without loss of generalization, we assume \struct interprets the neccessary variables.
\begin{enumerate}
 \item For each term $f(\ttt)$ that occurs in \theory, it holds that $f(\ttt)^{\locdomsym{\pi}{A}(\struct)}=\left\{ \begin{array}{ll} \pi(f(\ttt)^\struct) & \text{ if $f|0 \in A$}\\
                                                                                              f(\ttt)^\struct &\text{ otherwise}
                                                                                             \end{array} \right.$
\item For each atom $a$ of the form $P(t_1,\ldots,t_n)$ or of the form $t_1=t_2$ that occurs in \theory, it holds that $a^{\locdomsym{\pi}{A}(\struct)} = a^\struct$. 
\item For each formula $\varphi$  that occurs in \theory,  $\varphi^{\locdomsym{\pi}{A}(\struct)} = \varphi^\struct$. 
\end{enumerate}
The first claim is proven by induction on the subterm relation. The induction step follows from the fact that $A$ is connectively closed. 
The second claim follows from the first, also using the fact that $A$ is connectively closed.
Consider for instance the case of atom $f(\ttt)=g(\ttt')$ occurring in \theory, with $f|0\in A$.
Then $g|0\in A$ (since $A$ is connectively closed), so $f(\ttt)^{\locdomsym{\pi}{A}(\struct)}=g(\ttt')^{\locdomsym{\pi}{A}(\struct)}$ iff $\pi(f(\ttt)^\struct)=\pi(g(\ttt')^\struct)$ iff $f(\ttt)^\struct=g(\ttt')^\struct$ (since $\pi$ is a permutation). 
The other cases are analogous.

The last claim follows by induction on the subformula relation since the value of a first-order formula is entirely determined by the value of the atoms occuring in it.
Consider for instance the case of formula $\exists x\colon \varphi$ occurring in \theory with $\argpos{x}{0} \in A$.
$(\exists x\colon \varphi)^\struct$ holds iff there exists a $d \in \domain$ such that $\varphi^{\struct[x:d]}$ holds.
By the induction hypothesis, $\varphi^{\struct[x:d]}=\varphi^{\locdomsym{\pi}{A}(\struct[x:d])}=\varphi^{\locdomsym{\pi}{A}(\struct)[x:\pi(d)]}$ (since $\argpos{x}{0} \in A$).
$(\exists x\colon \varphi)^{\locdomsym{\pi}{A}(\struct)}$ holds iff there exists a $d' \in \domain$ such that $\varphi^{\locdomsym{\pi}{A}(\struct)[x:d']}$ holds.
Without loss of generalization, let $d'=\pi(d)$, then $(\exists x\colon \varphi)^\struct=(\exists x\colon \varphi)^{\locdomsym{\pi}{A}(\struct)}$.
The other cases are analogous.
\end{proof} 

\noindent
\textit{Theorem~\ref{thm:locdomsymconditionmx}}\\
\noindent
Let $MX(\theory,\structin)$ be a model expansion problem with decomposition $MX(\theorystar,\structinstar)$.
If $A$ is connectively closed under $\theorystar$ and $\locdomsym{\pi}{A}(\structinstar)=\structinstar$
then $\locdomsym{\pi}{A}$ is a symmetry for $MX(\theory,\structin)$.
\begin{proof}
Since $MX(\theory,\structin)$ and $MX(\theorystar,\structinstar)$ have the same set of solutions, it suffices to prove that $\locdomsym{\pi}{A}$ is a symmetry of $MX(\theorystar,\structinstar)$.
Firstly, due to the connectively closed condition, $\locdomsym{\pi}{A}$ is a symmetry of $\theorystar$, so $\structinstar \sqcup \structout \models \theorystar$ iff $\locdomsym{\pi}{A}(\structinstar \sqcup \structout) \models \theorystar$. 
Secondly, since $\locdomsym{\pi}{A}(\structinstar)=\structinstar$, $\structinstar \sqcup \structout \models \theorystar$ iff $\structinstar \sqcup \locdomsym{\pi}{A}(\structout) \models \theorystar$, so $\locdomsym{\pi}{A}$ is a symmetry for $MX(\theorystar,\structinstar)$.
\end{proof}

\noindent
\textit{Theorem~\ref{thm:complete_breaking}}\\
\noindent
Let $MX(\theory,\structin)$ be a model expansion problem, $\delta$
an $A$-interchangeable subdomain,
$\preceq_\domain$ a total order on domain \domain and $s(d)$ the successor of $d$ in $\delta$ according to $\preceq_\domain$.
If $A$ contains at most one argument position $\argpos{S}{i}$ for each symbol $S \in \vocout$, then the conjunction of lex-leader constraints
\[ \{ \lex{\locdomsym{(d~s(d))}{A}} \mid d\in \delta \} \]
is a complete symmetry breaking constraint for the subdomain
interchangeability group $\locdomintch{\delta}{A}$.
\begin{proof}
To prove this theorem, we (1) convert the task of finding a solution to a model expansion problem to a constraint programming problem, where an assignment over a set of Boolean variables $V$ has to be found.
Further, we show that (2) a subset of these Boolean variables can be organized as a matrix $M_\delta$, where each permutation over the rows of $M_\delta$ corresponds to a permutation over $\delta$.
The interchangeability group $\locdomintch{\delta}{A}$ then corresponds to a row interchangeability symmetry group induced by permuting $M_\delta$'s rows.
Using a result from constraint programming, such row interchangeability symmetry groups are broken completely by posting a lex-leader constraint (based on the appropriate row ordering) for each symmetry induced by the swap of two consecutive rows~\cite{row_column_sym_csp,cspsat/DevriendtBB14}. This corresponds to posting $\{ \lex{\locdomsym{(d~s(d))}{A}} \mid d\in \delta \}$, ending the proof.

(1) Given a vocabulary \vocout and a domain \domain, finding a $\vocout$-structure consists of deciding for each $\bar{d} \in D^n$ whether $\bar{d} \in S^{\structout}$ for each symbol $S/n \in \vocout$.
Hence, a model expansion problem $MX(\theory,\structin)$ can be seen as finding an assignment to a set of Boolean variables $V=\{S(\bar{d}) \mid S/n \in \vocout, \bar{d} \in D^n\}$ such that $\structin \sqcup \structout \models \theory$.
A local domain symmetry $\locdomsym{\pi}{A}$ for $MX(\theory,\structin)$ now corresponds to a \emph{variable symmetry}~\cite{row_column_sym_csp} mapping 
\[ S(d_1,\ldots,d_n) ~~~ \text{ to } ~~~ S(\localperm{}{S}{1}(d_1),\ldots, \localperm{}{S}{n}(d_n)) \]
where $\localperm{}{S}{i}(d) = \pi(d)$ if $\argpos{S}{i} \in A$ and $\localperm{}{S}{i}(d) =d$ otherwise.

(2) The variables in $V$ that are not fixed by some $\locdomsym{\pi}{A} \in \locdomintch{\delta}{A}$ are those $S(\ldots,d_{j-1},\delta_i,d_{j+1},\ldots)$ where $\delta_i \in \delta$ is the $j$th domain element of an $S$-tuple with $\argpos{S}{j} \in A$.
We can partition this subset into ``rows'' $R_{\delta_i} = \{S(\ldots,d_{j-1},\delta_i,d_{j+1},\ldots) \mid \argpos{S}{j} \in A, d_k \in \domain\}$ where $\delta_i$ is fixed.
It is clear that $\locdomsym{\pi}{A}(R_{\delta_i})=R_{\pi(\delta_i)}$, so a permutation of the set of rows corresponds to a symmetry of $\locdomintch{\delta}{A}$.
Since $A$ contains at most one argument position for each $S \in \vocout$, these rows are pairwise disjoint, and under some column organization form the requested matrix $M_\delta$.
\end{proof}

\newcommand\dA[1]{#1_{A}^+}
\newcommand\dnotA[1]{#1_{A}^-}
\noindent
\textit{Theorem~\ref{thm:domainpermutationgraph}}\\
\noindent
Let $\struct$ be a $\voc$-structure with domain $\domain$ and $A$ a set of argument positions.

There exists a bijection between the automorphism group of the domain permutation graph \dpg{\struct}{A} and the group of domain permutations $\pi$ such that $\locdomsym{\pi}{A}(\struct)=\struct$. This bijection maps an automorphism $\tau$ to domain permutation $\pi$ iff $\tau(d)=\pi(d)$ for all $DE$ vertices (equated with domain elements) $d$.

\begin{proof}
We prove the bijection by showing that all induced structure transformations $\locdomsym{\pi}{A}$ with $\locdomsym{\pi}{A}(\structin)=\structin$ correspond to an automorphism of \dpg{\structin}{A} ($\Rightarrow$) and vice versa ($\Leftarrow$).

First, some preliminaries. For a given symbol $S$, let each tuple $(d_1,\ldots,d_n) \in S^\structin$ be split as two tuples $\dA{d}\dnotA{d}$ such that $\dA{d}=\{d_i \mid \argpos{S}{i} \in A\}$ and $\dnotA{d}=\{d_i \mid \argpos{S}{i} \not \in A\}$.
Let $\pi$ naturally extend to tuples: $\pi((d_1,\ldots,d_n))=(\pi(d_1),\ldots,\pi(d_n))$.
The symmetrical interpretation $\locdomsym{\pi}{A}(\structin)$ can then be described as $\{\pi(\dA{d})\dnotA{d} \mid \dA{d}\dnotA{d} \in S^\structin\}$, so $\locdomsym{\pi}{A}(\structin)=\structin$ iff for all symbols $S$, $\dA{d}\dnotA{d} \in S^\structin$ iff $\pi(\dA{d})\dnotA{d} \in S^\structin$.
Also, without loss of generalization, let an IT vertex's label be $S(\dA{d}\dnotA{d})$ for symbol $S$.
Lastly, $(v,w) \in E$ denotes that graph $E$ has an (undirected) edge between vertices $v$ and $w$.


($\Rightarrow$)
If $\locdomsym{\pi}{A}(\structin)=\structin$, 
$\locdomsym{\pi}{A}$ corresponds to a permutation $\alpha$ of the vertices of \dpg{\structin}{A}: $\alpha(d)=\pi(d)$ (for DE vertices), $\alpha(\argnode{d}{i})=\argnode{\pi(d)}{i}$ (for AP vertices), $\alpha(S(\dA{d}\dnotA{d}))=S(\pi(\dA{d})\dnotA{d})$ (for IT vertices).
We show that $\alpha$ is an automorphism of \dpg{\structin}{A}.

By the definition of \dpg{\structin}{A}, $\alpha$ preserves the colors.
To show that $\alpha$ preserves the edges, we need to show that 
$(v,w) \in \dpg{\structin}{A}$ iff
$(v,w) \in \alpha(\dpg{\structin}{A})$.
Firstly, remark that $\alpha$ maps each vertex in a layer to another vertex in that layer, so we only need to check whether the edges between (1) DE-AP and (2) AP-IT are conserved.

(1) The following statements are equivalent
\begin{align*}
(\alpha(d),\alpha(\argnode{e}{i})) \in \alpha(\dpg{\structin}{A}) & \\
(d,\argnode{e}{i}) \in \dpg{\structin}{A} &\text{ ($\alpha$ is a permutation of vertices) } \\
d=e &\text{ (definition of domain permutation graph)  } \\
\pi(d)=\pi(e) &\text{ ($\pi$ is a permutation)  } \\
(\pi(d),\argnode{\pi(e)}{i}) \in \dpg{\structin}{A} &\text{ (definition of domain permutation graph) } \\
(\alpha(d),\alpha(\argnode{e}{i})) \in \dpg{\structin}{A} &\text{ (definition of $\alpha$) }
\end{align*}

(2) Similarly, the following statements are equivalent
\begin{align*}
(\alpha(\argnode{d}{i}),\alpha(S(\dA{d}\dnotA{d})) \in \alpha(\dpg{\structin}{A}) & \\
(\argnode{d}{i},S(\dA{d}\dnotA{d})) \in \dpg{\structin}{A} &\text{ ($\alpha$ is a permutation of vertices) } \\
d_i \in \dA{d} &\text{ (definition of domain permutation graph)  } \\
\pi(d_i) \in \pi(\dA{d}) &\text{ ($\pi$ is a permutation)  } \\
(\argnode{\pi(d)}{i},S(\pi(\dA{d})\dnotA{d})) \in \dpg{\structin}{A} &\text{ (definition of domain permutation graph) } \\
(\alpha(\argnode{d}{i}),\alpha(S(\dA{d}\dnotA{d})) \in \dpg{\structin}{A} &\text{ (definition of $\alpha$) }
\end{align*}

($\Leftarrow$)
We must show that an automorphism $\alpha$ of $\dpg{\structin}{A}$ corresponds to an $A,\pi$-induced structure transformation $\locdomsym{\pi}{A}$ such that $\locdomsym{\pi}{A}(\structin)=\structin$.

Notice that, since $\alpha$ is an automorphism of a three-layered graph with different colors for each layer DE, AP and IT, we can write it as a composition of three permutations $\alpha_{DE} \circ \alpha_{AP} \circ \alpha_{IT}$.
As there exists a bijection between DE and the domain \domain of \structin, we assume $\alpha_{DE}\backsimeq\pi$, with $\pi$ a permutation of \domain.

We now show that
(1) $\alpha(\argnode{d}{i})=\argnode{\pi(d)}{i}$ and
(2) $\alpha(S(\dA{d}\dnotA{d}))=S(\pi(\dA{d})\dnotA{d})$.
From this, it follows that $\alpha$ represents a structure transformation $\locdomsym{\pi}{A}$ mapping tuples $\dA{d}\dnotA{d}$ to $\pi(\dA{d})\dnotA{d}$, and hence, $\locdomsym{\pi}{A}(\structin)=\structin$.

(1) Since $\argnode{d}{i}$ and $\argnode{e}{j}$ have the same color iff $i=j$, $\alpha(\argnode{d}{i})=\argnode{e}{i}$ for some domain element $e$. As each vertex $\argnode{d}{i}$ is connected to exactly one vertex $d$, $\alpha(\argnode{d}{i})=\argnode{\pi(d)}{i}$.

(2) Since $S(\dA{d}\dnotA{d})$ and $R(\dA{e}\dnotA{e})$ have the same color iff $S=R$ and $\dnotA{d}=\dnotA{e}$, $\alpha(S(\dA{d}\dnotA{d}))=S(\dA{e}\dnotA{d})$ for some tuple domain elements $e$. 
All that is left to show is that $\dA{e}=\pi(\dA{d})$.
For this, note that $S(\dA{d}\dnotA{d})$ is connected only to $\argnode{d}{i}$ for each $d$ on index $i$ in $\dA{d}$. 
As $\alpha$ is an automorphism that maps $\argnode{d}{i}$ to $\argnode{\pi(d)}{i}$, $\alpha(S(\dA{d}\dnotA{d})$ must be connected only to all $\argnode{\pi(d)}{i}$.
The only vertex doing so (taking colors into account) is $S(\pi(\dA{d})\dnotA{d})$.
\end{proof}



\end{document}